\documentstyle[eqsecnum,prd,aps,epsf]{revtex}
\draft
\begin{document}

\thispagestyle{empty}
{\baselineskip0pt
\leftline{\baselineskip16pt\sl\vbox to0pt{\hbox
{\it Department of Physics}
               \hbox{\it Waseda University}\vss}}
\rightline{\baselineskip16pt\rm\vbox to20pt{\hbox{WU-AP/98/00}
	\hbox{OU-TAP 137}\hbox{OCU-PHYS-174}
\vss}}%
}
\vskip1cm
\begin{center}{\large \bf
Power, energy, and spectrum of a naked singularity explosion}
\end{center}
\vskip1cm
\begin{center}
 {\large 
Tomohiro Harada
\footnote{ Electronic address: harada@gravity.phys.waseda.ac.jp}}\\
{\em Department of Physics, Waseda University, Ohkubo, Shinjuku, Tokyo 169-8555, Japan}\\
{\large Hideo Iguchi
\footnote{ Electronic address: iguchi@vega.ess.sci.osaka-u.ac.jp}}\\
{\em Department of Earth and Space Science, Graduate School of Science,
 Osaka University,}\\
{\em Toyonaka, Osaka 560-0043, Japan} \\
{\large Ken-ichi Nakao
\footnote{ Electronic address: knakao@sci.osaka-cu.ac.jp}}\\
{\em Department of Physics,~Osaka City University,
Osaka 558-8585,~Japan}\\
\end{center}

\begin{abstract}
It is well known that a naked singularity occurs in the 
gravitational collapse of an inhomogeneous dust ball
from an initial density profile which is physically reasonable.
In this paper we show 
that explosive radiation is 
emitted during the formation process of the naked singularity
while we fix the background spacetime.
The energy flux is proportional to $(t_{\rm CH}-t)^{-3/2}$
for a minimally coupled massless scalar field, while it is 
proportional to $(t_{\rm CH}-t)^{-1}$ for a
conformally coupled massless scalar field, where 
$t_{\rm CH}-t$ is the ``remaineing time'' until the distant observer 
could observe the singularity 
if the naked singularity was formed.
As a consequence, the radiated energy grows unboundedly
for both scalar fields.
The amount of the power and 
energy depends on the parameters which characterize
the initial density profile but do not depend
on the gravitational mass of the cloud.
In particular, there is a characteristic frequency $\nu_{s}$ 
of the singularity above which 
the divergent energy is radiated.
The energy flux is dominated by particles of which
the wavelength is about $t_{\rm CH}-t$ at each moment.
The observed total spectrum is nonthermal, 
i.e., $\nu dN/d\nu \sim (\nu/\nu_{s})^{-1}$ for $\nu>\nu_{s}$.
If the naked singularity formation could continue 
until a considerable fraction 
of the total energy of the dust cloud is radiated,
the radiated energy would reach about 
$10^{54}(M/M_{\odot})\mbox{erg}$.   
The calculations are based on the 
geometrical optics approximation which turns out to be 
consistent for a rough order estimate. 
The analysis does not depend on whether or not
the naked singularity occurs in its exact meaning.
This phenomenon may provide a new candidate for a source of
ultrahigh energy cosmic rays or a central engine of
$\gamma$-ray bursts.
 
\end{abstract}


\section{introduction}
\label{sec:introduction}
It was shown that the collapse of an inhomogeneous dust ball
results in shell-focusing naked singularity 
formation~\cite{es1979,christodoulou1984,newman1986,jd1993}.
Ori and Piran~\cite{op1987,op1988,op1990} 
and Harada~\cite{harada1998}
showed that naked singularity 
formation is possible from 
the spherical collapse of a perfect 
fluid with a very soft equation of state.
Moreover, a kind of runaway collapse
in Newtonian gravity looks like naked singularity 
formation in general relativity.
These strongly suggest that the collapse in the Universe
will lead to a drastic growth of spacetime curvature which
is not covered by the event horizon. 

In this respect, a number of researchers have examined 
the emission during naked singularity formation.
In classical theory,
Shapiro and Teukolsky~\cite{st1991,st1992} reported
the drastic growth of a curvature polynomial 
but no apparent horizon in 
their numerical simulation of the axisymmetric collapse
of a prolate spheroid of collisionless particles. 
They also reported that little gravitational radiation
was emitted in their simulation.
On the other hand, Nakamura, Shibata and Nakao~\cite{nsn1993}
suggested that the naked singularity formation 
in the collapse of a prolate spheroid
may provide a strong source of gravitational waves.
Recently, Iguchi, Nakao, and Harada~\cite{inh1998} and
Iguchi, Harada, and Nakao~\cite{ihn1999a,ihn1999b} examined 
the behavior of nonspherical linear perturbations of the collapse
of an inhomogeneous dust ball.
One of their main results is that 
perturbations of a curvature polynomial grow
unboundedly while the energy flux of the gravitational waves remains
finite as the Cauchy horizon is approached,
which implies that the instability is rather milder.

On the other hand, Hawking~\cite{hawking1975} showed that
thermal radiation is emitted in
the gravitational collapse to a black hole.
The emission can be interpreted as particle creation due to
the time variation of a strong spacetime curvature.
In the collapse to a black hole,
the magnitude of the spacetime curvature of 
the observable region is, at most, the order of 
the inverse square of its gravitational mass
because the singularity is covered by the event horizon.
On the other hand, in the collapse to naked singularity,
the magnitude of the spacetime curvature outside 
the event horizon grows unboundedly
and the strongly curved region can be seen by an observer.
This fact suggests that 
naked singularity formation may be a strong source
of emission due to the quantum effect.
In this context, Ford and Parker~\cite{fp1978}
calculated the radiation during the formation process of 
shell-crossing naked singularity. 
They found that both the power and the energy remain finite,
which is because of the weakness of 
the shell-crossing singularity.
They also showed that the collapse of 
supercritically charged matter to the naked singularity
emitts unbounded power and energy, although 
this spacetime is not a solution of the Einstein equation.
Hiscock, Williams and Eardley~\cite{hwe1982}
showed the diverging energy flux in the 
self-similar collapse of
a null dust and also calculated the expectation value of
stress-energy tensor in its two-dimensional version.
Barve et al.~\cite{bsvw1998a}
showed the diverging power
of radiation in the self-similar collapse of a dust ball.
They also calculated the expectation value of stress-energy tensor
in its two-dimensional version~\cite{bsvw1998b}.
Vaz and Witten~\cite{vw1998} also calculated 
the Bogoliubov coefficients which
relate to the spectrum of emission for this model.
The above studies were all based on 
the geometrical optics approximation.

The last two examples in which the diverging power is emitted 
are self-similar collapse.
However, the self-similar collapse is a particular solution 
among gravitational
collapse solutions in general relativity.
Moreover, we do not know whether or not the central part of the
realistic collapse of the nonzero pressure perfect fluid tends 
to be self similar in a strong-gravity regime 
such as a shell-focusing
singularity, although such tendency has been 
observed in the Newtonian regime.
For the self-similar collapse of a dust ball,
it has been shown that the redshift at the center 
becomes infinite and
that the curvature strength of the naked singularity 
is very strong.
Moreover, it is known that such a solution 
does not allow the initial density profile which is a $C^{\infty}$
function with respect to the local Cartesian coordinates.
On the other hand,
Christodoulou~\cite{christodoulou1984} imposed 
the $C^{\infty}$ condition 
on the initial density distribution in his famous paper on the 
violation of cosmic censorship in the collapse of a dust ball.
For the $C^{\infty}$ case, the redshift is finite and
the naked singularity is not very strong.
See~\cite{djd1999,hni1999,dwivedi1998} for details.
Though the Einstein equation does not require
such strong differentiability to initial data,
we usually set such initial data in most 
astrophysical numerical simulations.
Although this model is shown to be unstable
to nonspherical perturbations, 
the model is still good for the sufficiently spherically symmetric 
collapse of dust until the nonspherical perturbations grow sufficiently.
It will be important to examine whether or not 
the divergence of the radiated energy flux in the 
self-similar collapse only
comes from the uniqueness of self-similarity.
In this context,
the authors have already reported explosive radiation
during the formation process of the naked singularity
which occurs
in the collapse of an inhomogeneous dust ball 
from the initial density profile 
which is a $C^{\infty}$ function~\cite{hin2000}.
In this paper, we present the details of the calculations in the 
above paper and examine power, energy, and spectrum
of that radiation process.

This paper is organized as follows.
In Sec.~\ref{sec:particle_creation},
we will recall the least of quantum particle creation
by a collapsing ball based on the 
geometrical optics approximation.
In Sec.~\ref{sec:LTB}, we present the background solution
of spacetime which describes the 
collapse of an inhomogeneous dust ball.
In Sec.~\ref{sec:power},
we present numerical results of power and energy
due to the quantum particle creation
and give some physical discussions.
In Sec.~\ref{sec:spectrum}, we determine 
the spectrum of the radiation
which is important not only astrophysically but also
in estimating the validity of the geometrical optics approximation.
In Sec.~\ref{sec:null_geodesics}, we analytically examine
null geodesics and redshifts in this spacetime and thereby
understand some features of the numerical results.
In Sec.~\ref{sec:validity}, we estimate the 
validity of the geometrical optics approximation.
In Sec.~\ref{sec:discussions}, some implications of the
emission from the forming 
naked singularity are discussed.
In Sec.~\ref{sec:conclusions}, we conclude the paper.
We use the units in which $G=c=\hbar=1$ unless
explicitly stated.
We follow the sign conventions of Misner, Thorne, and 
Wheeler~\cite{mtw1973}. 

\section{particle creation by a collapsing ball}
\label{sec:particle_creation}
We consider both minimally and conformally coupled 
massless scalar fields in a four-dimensional spacetime
which is spherically symmetric and asymptotically flat. 
Let $T,R,\theta,\phi$ denote the usual 
quasi-Minkowskian time and spherical coordinates,
which are asymptotically related to null coordinates 
$u$ and $v$ by $u\approx T-R$ and $v\approx T+R$.
An incoming null ray $v=\mbox{const}$, originating on ${\Im}^{-}$,
propagates through the geometry becoming an outgoing null ray 
$u=\mbox{const}$, and arriving on ${\Im}^{+}$ at a value $u=F(v)$.
Conversely, we can trace a null ray from $u$ on ${\Im}^{+}$ to
$v=G(u)$ on ${\Im}^{-}$, where $G$ is the inverse of $F$.
Here, we assume that the geometrical optics approximation is valid.
The validity of this approximation will be discussed later.
The geometrical optics approximation 
implies that the trajectories of the null rays
give a surface of the constant phase.
Then, in the asymptotic region, the mode function 
which contains an ingoing mode of the 
standard form on ${\Im}^{-}$ is the following:
\begin{equation}
u^{in}_{\omega l m}\approx \frac{1}
{\sqrt{4\pi \omega} R}(e^{-i\omega v}-e^{-i
\omega G(u)})Y_{lm}(\theta,\phi),
\end{equation}
where we have imposed the 
reflection symmetry condition at the center.
On the other hand, in the asymptotic region, the mode function 
which contains an outgoing mode of the 
standard form on ${\Im}^{+}$ is the following:
\begin{equation}
u^{out}_{\omega l m}\approx \frac{1}
{\sqrt{4\pi \omega} R}(e^{-i\omega 
F(v)}-e^{-i\omega u})Y_{lm}(\theta,\phi).
\end{equation}
Note that in the above we have normalized the mode functions as
\begin{equation}
(u_{\omega l m},u_{\omega^{\prime} l^{\prime} m^{\prime}})
=\delta(\omega-\omega^{\prime})\delta_{ll^{\prime}}\delta_{mm^{\prime}},
\end{equation}
where the inner product is defined by integration on
the spacelike hypersurface $\Sigma$ as
\begin{equation}
(\psi,\chi)=-i\int_{\Sigma}(\psi\chi^{*}_{,\mu}-\psi_{,\mu}
\chi^{*})\sqrt{g_{\Sigma}}d\Sigma^{\mu}.
\end{equation}

Using the above mode functions we can express the scalar field $\phi$ as
\begin{eqnarray}
\phi&=&\sum_{l,m}\int d\omega^{\prime}
(\mbox{\boldmath{$a$}}^{in}_{\omega^{\prime}lm}
u^{in}_{\omega^{\prime}lm}+
\mbox{\boldmath{$a$}}^{in\dagger}_{\omega^{\prime}lm}
u^{in*}_{\omega^{\prime}lm}), \\
\phi&=&\sum_{l,m}\int 
d\omega^{\prime}(\mbox{\boldmath{$a$}}^{out}_{\omega^{\prime}lm}
u^{out}_{\omega^{\prime}lm}+
\mbox{\boldmath{$a$}}^{out\dagger}_{\omega^{\prime}lm}
u^{out*}_{\omega^{\prime}lm}).
\end{eqnarray}
According to the usual procedure of canonical quantization, 
we obtain the following commutation relations
\begin{eqnarray}
\left[\mbox{\boldmath{$a$}}_{\omega lm}^{in}, 
\mbox{\boldmath{$a$}}_{\omega^{\prime}l^{\prime}m^{\prime}}
^{in\dagger}\right]&=&
\delta(\omega-\omega^{\prime})
\delta_{l l^{\prime}}\delta_{m m^{\prime}}, \\
\left[
\mbox{\boldmath{$a$}}_{\omega lm}^{out}, 
\mbox{\boldmath{$a$}}_{\omega^{\prime}l^{\prime}m^{\prime}}
^{out \dagger}\right]&=&
\delta(\omega-\omega^{\prime})
\delta_{l l^{\prime}}\delta_{m m^{\prime}},
\end{eqnarray}
where it is noted that the Lagrangian in the Minkowski spacetime
is common for both minimally 
and conformally coupled scalar fields.
Here, $\mbox{\boldmath{$a$}}_{\omega lm}^{in}$ 
and $\mbox{\boldmath{$a$}}_{\omega lm}^{out}$ 
are interpreted as annihilation operators 
corresponding to in and out modes, respectively.
Then we set the initial quantum state to in vacuum, i.e.,
\begin{equation}
  \mbox{\boldmath{$a$}}_{\omega lm}^{in}|0>=0.
\end{equation}

The power for fixed $l$ and $m$ is given by
estimating the expectation value of the stress-energy tensor
through the point-splitting regularization in a flat spacetime
as~\cite{fp1978}
\begin{equation}        
P_{lm}\equiv \int <T_{T}^{~R}>R^2d\Omega
=\frac{1}{24\pi}\left[\frac{3}{2}\left(
\frac{G^{\prime\prime}}{G^{\prime}}\right)^2-
\frac{G^{\prime\prime\prime}}{G^{\prime}}\right]
=\frac{1}{48\pi}
\left(\frac{G^{\prime\prime}}{G^{\prime}}\right)^2
-\frac{1}{24\pi}\left(\frac{G^{\prime\prime}}
{G^{\prime}}\right)^{\prime}
\label{eq:llm}
\end{equation}
for a minimally coupled scalar field, and
\begin{equation}        
\hat{P}_{lm}\equiv \int <\hat{T}_{T}^{~R}>
R^2d\Omega=\frac{1}{48\pi}
\left(\frac{G^{\prime\prime}}{G^{\prime}}\right)^2
\label{eq:hatllm}
\end{equation}  
for a conformally coupled scalar field.
It implies that the power depends on
the method of coupling of the scalar field with 
gravity.
However, if 
\begin{equation}
\left.\frac{G^{\prime\prime}}{G^{\prime}}\right|_{a}
=\left.\frac{G^{\prime\prime}}{G^{\prime}}\right|_{b}
\end{equation}
holds,
the radiated energy from $u=a$ to
$u=b$ of a minimally coupled field
\begin{equation}
E_{lm}\equiv \int_{a}^{b}P_{lm}du
\end{equation}
and that of a conformally coupled field
\begin{equation}
\hat{E}_{lm}\equiv \int_{a}^{b}\hat{P}_{lm}du
\end{equation}
coincide exactly. 
The total power is given by summation of all $(l,m)$.
The simple summation diverges.
This is because we have neglected the back scattering effect by
the curvature potential, which will reduce the radiated flux
considerably for larger $l$.
Therefore we should recognize that the above expressions of 
the power, Eqs.
(\ref{eq:llm}) and (\ref{eq:hatllm}),
are a good approximation only for smaller $l$.
Hereafter we omit the suffix $l$ and $m$.

The spectrum of radiation is derived 
from the Bogoliubov coefficients
which relate the in and out modes given by~\cite{bd1984}
\begin{eqnarray}
\alpha_{\omega^{\prime}\omega}&=&(u^{in}_{\omega^{\prime}},
u^{out}_{\omega})=\frac{1}{2\pi}
\sqrt{\frac{\omega^{\prime}}{\omega}}\int^{\infty}_{-\infty}
dv e^{i\omega F(v)-i\omega^{\prime}v}, 
\label{eq:alpha}\\
\beta_{\omega^{\prime}\omega}&=&
-(u^{in}_{\omega^{\prime}},u^{out*}_{\omega})=
-\frac{1}{2\pi}
\sqrt{\frac{\omega^{\prime}}{\omega}}\int^{\infty}_{-\infty}
dv e^{-i\omega F(v)-i\omega^{\prime}v}.
\label{eq:beta}
\end{eqnarray}
From the above equations, we find the following useful relation:
\begin{equation}
\beta_{\omega^{\prime}\omega}=-i\alpha_{\omega^{\prime}(-\omega)}.
\end{equation}
The Bogoliubov coefficients satisfy the following relations:
\begin{eqnarray}
\int^{\infty}_{0}d\tilde{\omega}(\alpha_{\tilde{\omega}
\omega}
\alpha^{*}_{\tilde{\omega}\omega^{\prime}}
-\beta_{\tilde{\omega}\omega}
\beta^{*}_{\tilde{\omega}\omega^{\prime}})&=&
\delta(\omega-\omega^{\prime}), 
\label{eq:bogo} \\
\int^{\infty}_{0}d\tilde{\omega}(\alpha_{\tilde{\omega}\omega}
\beta_{\tilde{\omega}\omega^{\prime}}
-\beta_{\tilde{\omega}\omega}
\alpha_{\tilde{\omega}\omega^{\prime}})&=&0.
\end{eqnarray}
The expectation value $N(\omega)$ of the 
particle number of frequency $\omega$
on ${\Im}^{+}$ is obtained by
\begin{equation}
N(\omega)
=\int^{\infty}_{0}d\omega^{\prime}|\beta_{\omega^{\prime}
\omega}|^2.
\label{eq:nomega}
\end{equation}

It is noted that these results are free of 
ambiguity coming from local curvature
because the regularization is done
in a flat spacetime.

\section{Lema\^{\i}tre-Tolman-Bondi solution}
\label{sec:LTB}
The spherically symmetric collapse of a dust fluid is
exactly solved~\cite{tolman1934,bondi1947}.
The solution is called the Lema\^{\i}tre-Tolman-Bondi (LTB)
solution.
The metric is given by
\begin{equation}
ds^2=-dt^2+\frac{(R_{,r})^{ 2}(t,r)}{1+f(r)}dr^2
+R^2(t,r)(d\theta^2+\sin^2\theta
d\phi^2)
\label{eq:metric}
\end{equation}
in the synchronous comoving coordinates.
The matter field is given by
\begin{eqnarray}
T^{\mu\nu}&=&\epsilon u^{\mu}u^{\nu}, \\
\epsilon &=&\frac{F^{\prime}(r)}{8\pi R^2 R_{,r}}.
\label{eq:epsilon}
\end{eqnarray}
$F(r)$ and $f(r)$ are arbitrary functions.
$F(r)$ is the mass function, which is twice the 
Misner-Sharp mass.
$f(r)$ is the energy function.
The areal radius $R$ satisfies the following equation:
\begin{equation}
(R_{,t})^2=\frac{F(r)}{R}+f(r).
\label{eq:energy}
\end{equation}
Equation~(\ref{eq:energy}) is integrated to the following form
\begin{equation}
t-t_{s}(r)=-\sqrt{\frac{R^3}{F}} g\left(-\frac{fR}{F}\right),
\end{equation}
where $g(y)$ is a positive function given by
\begin{equation}
g(y)= \left\{
\begin{array}{ll}
\displaystyle{\frac{\mbox{Arcsin}\sqrt{y}}{y^{3/2}}}-
\displaystyle{\frac{\sqrt{1-y}}{y}}&\qquad \mbox{for}~~0<y\le 1, \\
\displaystyle{\frac{2}{3}}&\qquad \mbox{for}~~y=0, \\
\displaystyle{\frac{-\mbox{Arcsinh}\sqrt{-y}}{(-y)^{3/2}}}-
\displaystyle{\frac{\sqrt{1-y}}{y}}&\qquad \mbox{for}~~y<0,
\end{array}\right.
\end{equation}
$t_s(r)$ is a constant of integration,
and the sign is chosen so that the collapsing phase is relevant.
Note that the metric has physical meaning only 
on $0\le r $ and $t<t_{s}(r)$.
We rescale the radial coordinate $r$ as follows
\begin{equation}
R(0,r)=r.
\end{equation}
From Eq.~(\ref{eq:epsilon}), if we require 
that the initial density profile
at $t=0$
is a $C^{\infty}$ function with 
respect to the local Cartesian coordinates,
$F(r)$ should be expanded around $r=0$ as
\begin{equation}
F(r)=F_{3}r^3+F_{5}r^5+F_{7}r^7+\cdots,
\label{eq:expansion}
\end{equation}
where we assume that $F_{3}$ is positive.
Then, $t_{s}(r)$ is given by
\begin{equation}
t_{s}(r)=\sqrt{\frac{r^3}{F}}g\left(-\frac{fr}{F}\right).
\end{equation}
We can easily find that
the time of the occurrence of shell-focusing singularity is given 
by $t_{s}(r)$, while the time of 
the occurrence of apparent horizon is
given by 
\begin{equation}
t_{ah}(r)=t_{s}(r)-Fg(-f).
\end{equation}
Hereafter we assume marginally bound collapse ($f=0$) for simplicity.

At an arbitrary radius $r$,
the LTB spacetime can be matched with the Schwarzschild spacetime 
\begin{equation}
ds^2=-\left(1-\frac{2M}{R}\right)dT^2+\left(1-\frac{2M}{R}\right)^{-1}
dR^2+R^2(d\theta^2+\sin^2\theta d\phi^2).
\end{equation}
The matching conditions with the Schwarzschild spacetime are obtained 
by requiring that the surface moves on the timelike geodesic and that
the three metric on the timelike hypersurface
should be continuous. The result is
\begin{eqnarray}
T&=&t-\frac{2r^{3/2}}{3\sqrt{2M}}-2\sqrt{2MR}
+2M\ln\frac{\sqrt{R}+\sqrt{2M}}
{\sqrt{R}-\sqrt{2M}}, 
\label{eq:match1}\\
R&=&R(t,r),
\label{eq:match2}\\
M&=&\frac{F(r)}{2},
\label{eq:match3}
\end{eqnarray}
at the surface $r=r_{sf}$.
The retarded time $u$ and the advanced time $v$ in the Schwarzschild 
spacetime are defined by
\begin{eqnarray}
u&=&T-R-2M\ln\frac{R-2M}{2M}, 
\label{eq:u}\\
v&=&T+R+2M\ln\frac{R-2M}{2M}.
\label{eq:v}
\end{eqnarray}

For the case of marginally bound collapse, $R$ is given by
\begin{equation}
R=r \left(1-\frac{3}{2}\sqrt{\frac{F}{r^3}} t\right)^{2/3}.
\label{eq:R}
\end{equation}
The occurrence of singularity is at the time 
\begin{equation}
t_{s}(r)=\frac{2}{3}\sqrt{\frac{r^3}{F}}.
\end{equation}
Each mass shell is trapped at the time
\begin{equation}
t_{ah}(r)=\frac{2}{3}\sqrt{\frac{r^3}{F}}-\frac{2}{3}F.
\end{equation}
We denote the time of the occurrence of the singularity
at the center as $t_{0}\equiv t_{s}(0)$.

Detailed analysis on 
the occurrence of naked shell-focusing singularity
has been done 
in~\cite{es1979,christodoulou1984,newman1986,jd1993,sj1996,jjs1996,jj1997}.
Here we present one of the results.
If we assume marginally bound collapse 
and assume the initial density profile expressed 
by Eq.~(\ref{eq:expansion}),
then the central shell-focusing singularity is naked 
if and only if $F_{5}$ is negative.
The singularity satisfies not the strong curvature condition
(SCC) defined 
by Tipler~\cite{tipler1977} but only the limiting focusing condition (LFC)
defined by Kr\'olak~\cite{krolak1987} 
for radial null geodesics which terminate
at the singularity.
On the other hand, the singularity satisfies
both LFC and SCC for radial timelike geodesics~\cite{djd1999}.

\section{power and energy}
\label{sec:power}
\subsection{Globally naked singularity}
\label{subsec:naked_singularity}
We can find the function $G$ or $F$ by solving the trajectories 
of outgoing and ingoing null rays in the dust cloud and
determining the retarded time $u$ and the advanced time $v$
through Eqs.~(\ref{eq:u}) and (\ref{eq:v})
at the time when the outgoing and ingoing null rays reach 
the surface boundary, respectively.
The trajectories of null rays in the dust cloud are given by
the following ordinary differential equation:
\begin{equation}
\frac{dt}{dr}=\pm R_{,r},
\label{eq:dtdr}
\end{equation} 
where the upper and lower signs denote
outgoing and ingoing null rays, respectively. 

We have solved the ordinary differential equation (\ref{eq:dtdr})
numerically.
We have used the Runge-Kutta method of the fourth order.
We have executed the quadruple precision calculations. 

\begin{figure}[htbp]
        \centerline{\epsfxsize 18cm \epsfysize 14cm \epsfbox{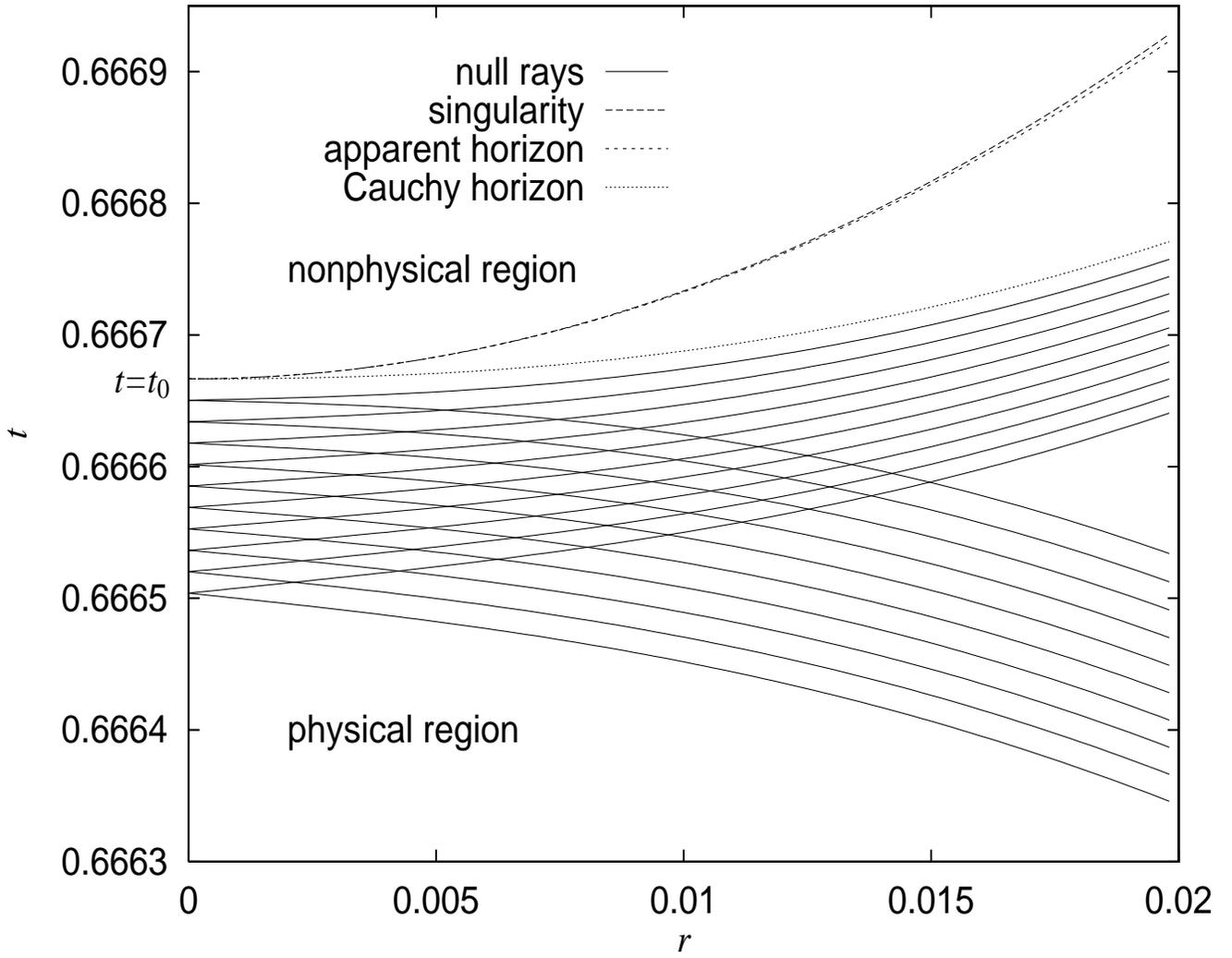}}
        \caption{Null rays interior of the 
dust cloud in the LTB spacetime.}
        \label{fg:rays_ltb}
\end{figure}

We choose the mass function as
\begin{equation}  
        F(r)=F_{3}r^3+F_{5}r^5.
\label{eq:mass}
\end{equation}
We find that the central singularity is globally naked for very 
small $r_{sf}$ if we fix the values of $F_{3}$ and $F_{5}$. 
Although we have calculated several models,
we only display the numerical results for the model with
$F_{3}=1$, $F_{5}=-2$ and $r_{sf}=0.02$ in an arbitrary unit
because the features are the same if there exists globally
naked singularity in the model.
The total gravitational mass $M$ is given by $M=3.9968\times 10^{-6}$
for this model. 
See Fig.~\ref{fg:rays_ltb} for trajectories of null geodesics.
We also indicate the location of the singularity, the apparent horizon,
and the Cauchy horizon in this figure.

\begin{figure}
        \centerline{\epsfxsize 18cm \epsfysize 14cm \epsfbox{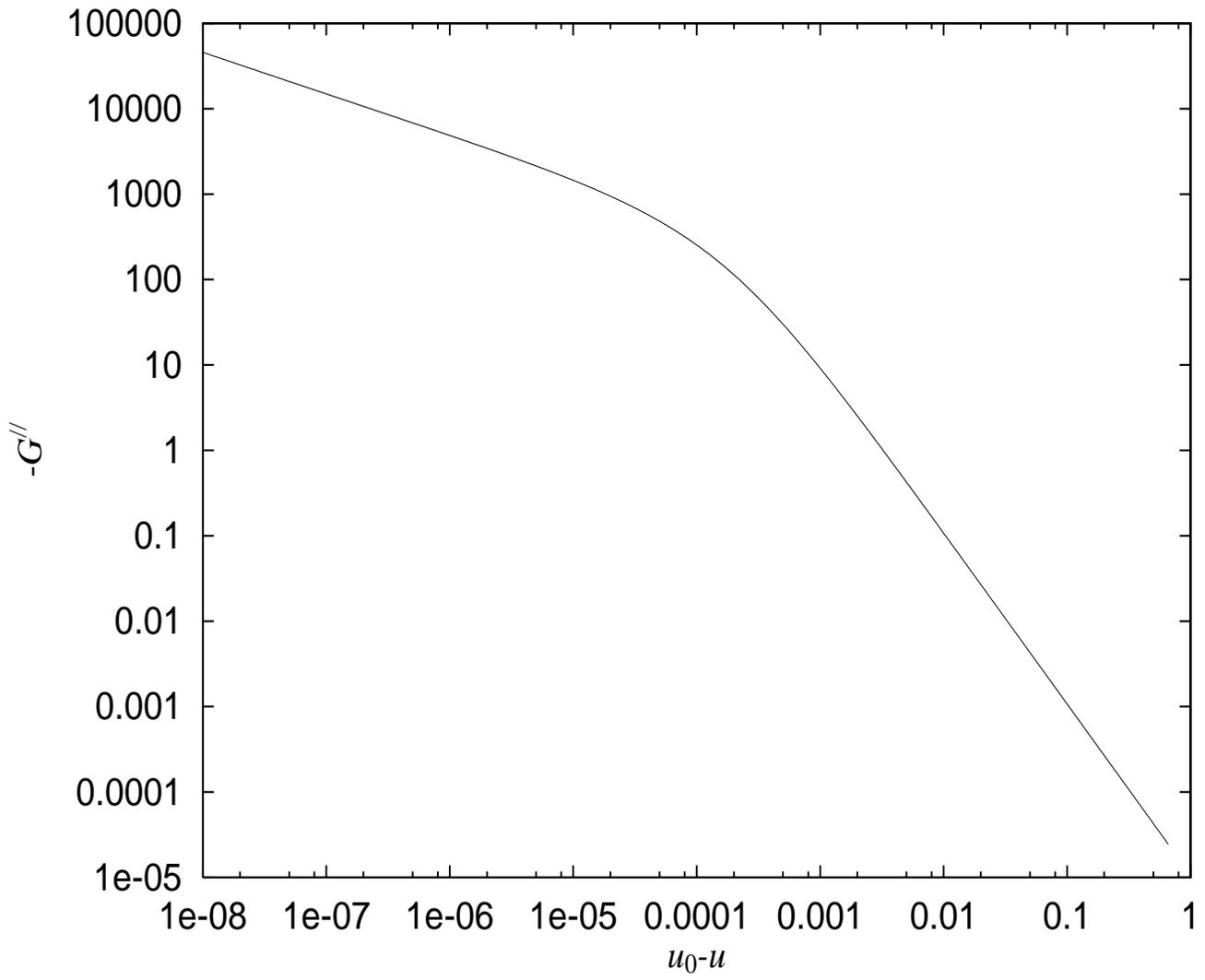}}
~(c)
        \caption{Plots of (a) $G(u)$, and (b) $G^{\prime}(u)$, (c)
 $G^{\prime\prime}(u)$ for the LTB spacetime.}
\label{fg:g_ltb}
\end{figure}

See Fig.~\ref{fg:g_ltb}. In Fig.~\ref{fg:g_ltb}(a), 
the relation between $u$ and $v$, i.e., the function $G(u)$
is shown, where 
$u_{0}$ ($v_{0}$) is defined as the retarded (advanced) 
time of the earliest 
light ray which originates from (terminates at) the singularity.
In Fig.~\ref{fg:g_ltb}(b), the first derivative $G^{\prime}(u)$ is shown.
This implies that $G^{\prime}(u)$ does not diverge
but converges to some positive constant $A$ with $0<A<1$.
In Fig.~\ref{fg:g_ltb}(c), it is found that
the second derivative $G^{\prime\prime}(u)$ does diverge 
as $u\to u_{0}$.
This figure shows that the behaviors of the 
growth of $G^{\prime\prime}(u)$ 
are different from each other during
early times ($10^{-4}\alt u_{0}-u $) and during late times 
($0 < u_{0}-u \alt 10^{-4}$).
During early times, the dependence on $u$ is written as
\begin{equation}
        G^{\prime\prime}\propto -(-u)^{-2},
        \label{eq:G2early}
\end{equation}
while, during late times, the dependence is written as
\begin{equation}
        G^{\prime\prime}\propto -(u_{0}-u)^{-1/2}.
        \label{eq:G2late}
\end{equation}

Here we determine the magnitude of the power by physical 
discussions.
We assume that the particle creation during early times is due to
the collapse of the dust cloud as a whole while that during late times
is due to the growth of the central curvature.
Then, we consider a constant of proportion which will appear
in Eqs.~(\ref{eq:G2early})
and (\ref{eq:G2late}).
First we should note that the coefficient must be
written by initial data because it must not depend on time.
Next we should note that we can identify any $t=\mbox{const}<t_{0}$ 
hypersurface with the initial hypersurface.
Since the constant of proportion directly enters 
the expression of the power which is physically meaningful,
the coefficient must be independent of
the choice of an initial slice.
On the other hand, Eqs.~(\ref{eq:G2early}) and (\ref{eq:G2late})
demand a dimensionful constant of proportion.
For early times, the only possible quantity is 
the gravitational mass $M$ of the dust cloud.
On the other hand, for late times, as derived 
in Appendix~\ref{sec:conserved_quantity},
the only possible quantity
is $(t_{0}^7 /l_{0}^{6})$, where $t_{0}$ is given by
\begin{equation}
t_{0}=\frac{2}{3}F_{3}^{-1/2},
\end{equation}
and $l_{0}$ denotes the
scale of inhomogeneity defined as
\begin{equation}
l_{0}\equiv\left(\frac{-F_{5}}{F_{3}}\right)^{-1/2}.
\end{equation}
Thus we can determine the coefficient except for a numerical factor as
\begin{equation}
        G^{\prime\prime}\approx -f_{e}M(-u)^{-2},
\end{equation}
for early times,
while the dependence is written as
\begin{equation}
G^{\prime\prime}\approx -f_{l}A\left(\frac{t_{0}^{7}}{l_{0}^{6}}
\right)^{-1/2}(u_{0}-u)^{-1/2},
\end{equation}
for late times,
where $f_{e}$ and $f_{l}$ are dimensionless 
positive constants of order unity.
It implies that there is a characteristic frequency of singularity
which is defined as
\begin{equation}
        \omega_{s}\equiv\frac{l_{0}^{6}}{t_{0}^7}.
        \label{eq:omegas}
\end{equation}
The early time behavior derived above is good for $-u\gg M$,
while the late time behavior is good for $0<u_{0}-u\ll \omega_{s}^{-1}$.
The turning point from the early time behavior 
to the late time behavior is roughly
estimated as
\begin{equation}
u_{0}-u\approx (M\omega_{s})^{2/3}\omega_{s}^{-1}=(M\omega_{s})^{-1/3}M.
\end{equation}
Here we define the turning point frequency $\omega_{tp}$ as
\begin{equation}
\omega_{tp}\equiv (M\omega_{s})^{-2/3}\omega_{s}=(M\omega_{s})^{1/3}M^{-1}.
\end{equation}
The above estimates show a good agreement with the numerical results. 
        
\begin{figure}[htbp]
        \centerline{\epsfxsize 18cm \epsfysize 14cm \epsfbox{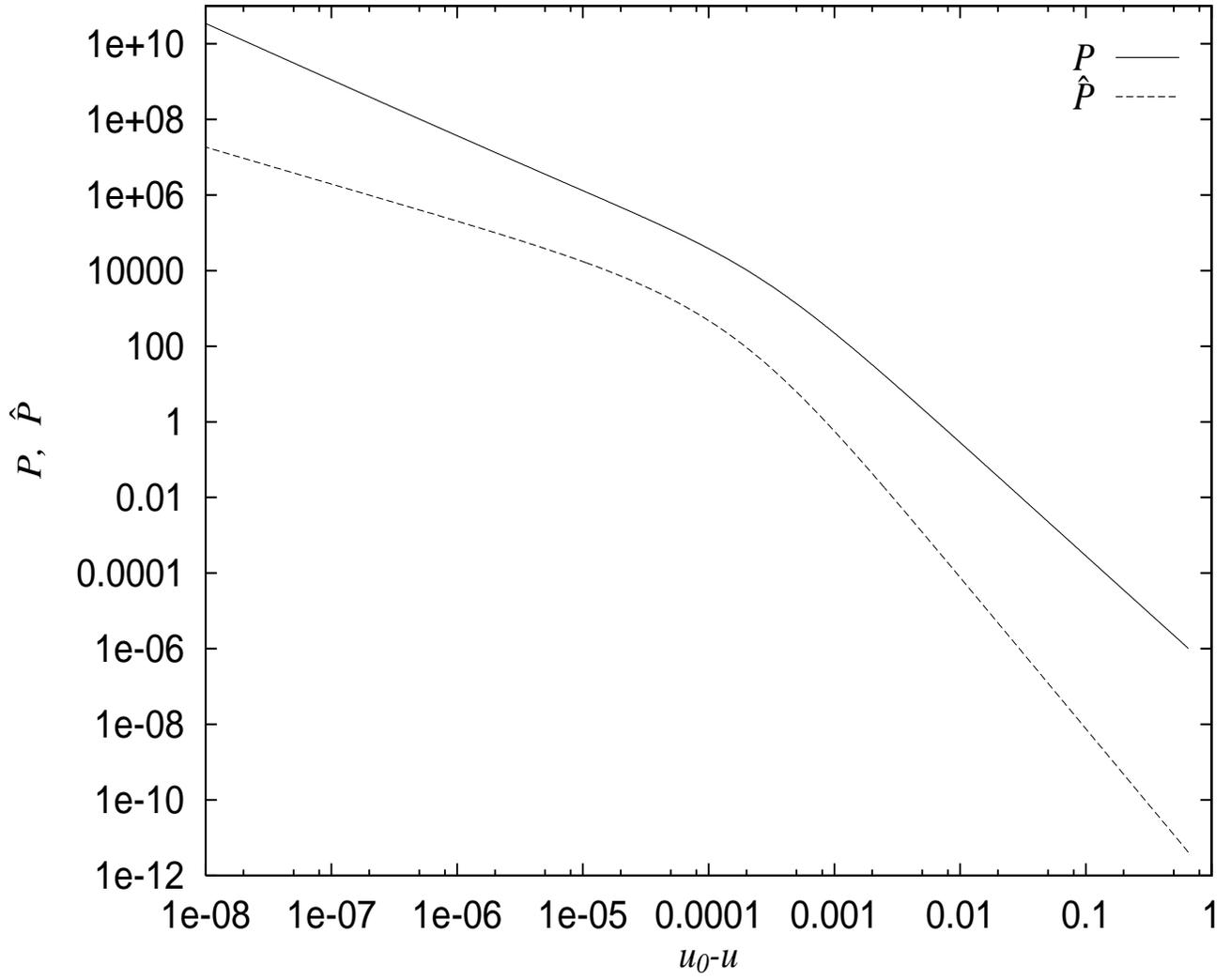}}
        \caption{Power for minimally and conformally
        coupled scalar fields in the LTB spacetime}
        \label{fg:p_ltb}
\end{figure}

Since $G^{\prime\prime}$ and therefore $G^{\prime\prime\prime}$ diverge 
as $u\to u_{0}$, the power of radiation diverges for 
both minimally and conformally coupled scalar fields.
For this model, we have numerically calculated 
the power of radiation by 
Eqs.~(\ref{eq:llm}) and (\ref{eq:hatllm}).
The results are displayed in Fig.~\ref{fg:p_ltb}.

Based on the above physical discussions, we can obtain
the formula for the power by the particle creation
using Eqs.~(\ref{eq:llm}) and (\ref{eq:hatllm}).
As seen in Fig.~\ref{fg:g_ltb}(b), we find  
\begin{equation}
        G^{\prime}\approx 1
\end{equation}
during early times, and
\begin{equation}
        G^{\prime}\approx A,
\end{equation}
during late times.
Then, the power during early times is obtained as
\begin{eqnarray}
P&\approx& \frac{1}{12\pi}f_{e}M(-u)^{-3}, \\
\hat{P}&\approx& \frac{1}{48\pi}f_{e}^2 M^2
(-u)^{-4}.
\end{eqnarray}
The power during late times is obtained as
\begin{eqnarray}
P&\approx& 
\frac{1}{48\pi}f_{l}
\omega_{s}^{1/2}(u_{0}-u)^{-3/2}, \\
\hat{P}&\approx&
\frac{1}{48\pi}f_{l}^2 
\omega_{s}(u_{0}-u)^{-1}.
\label{eq:hatl}
\end{eqnarray}
Therefore, the power diverges to positive infinity 
for both minimally and 
conformally coupled scalar fields as $u\to u_{0}$.
The radiated energy is estimated by integrating the power with time.
During late times, the radiated energy is estimated as
\begin{eqnarray}
E&\approx&\frac{1}{24\pi}f_{l}\omega_{s}^{1/2}
(u_{0}-u)^{-1/2}, \\
\hat{E}&\approx&\frac{1}{48\pi}f_{l}^2
\omega_{s}\ln\frac{(M\omega_{s})^{2/3}}{\omega_{s}(u_{0}-u)}.
\end{eqnarray}
Therefore, the total radiated energy diverges to positive infinity
for both minimally and conformally 
coupled scalar fields as $u\to u_{0}$.
However, in realistic situations,
we may assume that the naked singularity formation 
is prevented by some mechanism and that the quantum particle creation
is ceased at the time $u_{0}-u\approx \Delta t$.
In other words, $G^{\prime\prime}/G^{\prime}$ 
tends to vanish for $u_{0}-u\alt 
\Delta t$.
In such situations, the total divergence term in the 
expression of the power
of a minimally coupled scalar field gives no contribution to the
total radiated energy.
Therefore, the total energy for a minimally coupled 
scalar field and that for a conformally coupled one 
coincide exactly, i.e.,
\begin{equation}
E=\hat{E}\approx \frac{1}{48\pi}f_{l}^{2}
\omega_{s}\ln \frac{(M\omega_{s})^{2/3}}{\omega_{s}\Delta t}.
\end{equation}
The relation of $\Delta t$
and the maximum central energy density 
$\epsilon_{c}^{max}$ 
of the dust cloud at that time is given by
\begin{equation}
\epsilon_{c}^{max} \sim (\Delta t)^{-2}.
\end{equation}

From the above numerical results and physical 
discussions, we obtain the 
following formula for the function $G(u)$ or $F(v)$: 
\begin{equation}
G(u)\approx u+f_{e}M\ln(-u)
\end{equation}
or
\begin{equation}
F(v)\approx v-f_{e} M\ln(-v)
\end{equation}
for early times, and
\begin{equation}
G(u)\approx A (u-u_{0}) -\frac{4}{3}
A f_{l}\omega_{s}^{1/2}
(u_{0}-u)^{3/2}+v_{0}
\label{eq:GlateLTB}
\end{equation}
or
\begin{equation}
F(v)\approx A^{-1}(v-v_{0})+\frac{4}{3}f_{l}
\omega_{s}^{1/2}[A^{-1}(v_{0}-v)]^{3/2}+u_{0}
\label{eq:FlateLTB}
\end{equation}
for late times.

\subsection{Black hole}
\label{subsec:black_hole}
For comparison and for a test of the numerical code, 
we have also calculated the function $G$ for the 
Oppenheimer-Snyder collapse. 
The model is given by setting $F_{3}=1$,
$F_{5}=0$, and $r_{sf}=0.2$ ($M=4\times 10^{-3}$) in
Eq.~(\ref{eq:mass}) in an arbitrary unit. 

\begin{figure}[htbp]
 \centerline{\epsfxsize 18cm \epsfysize 14cm \epsfbox{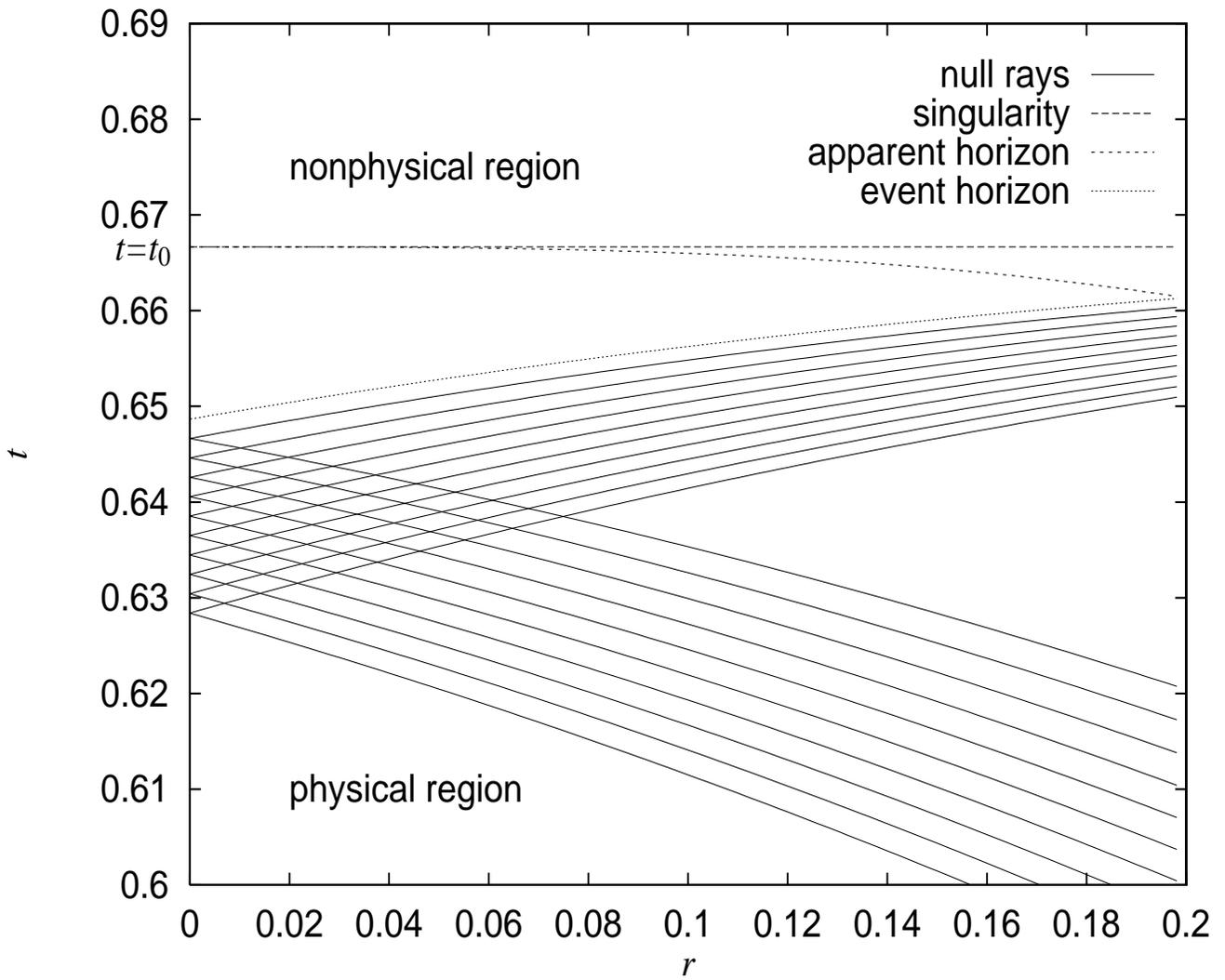}}
 \caption{Null rays interior of the dust cloud in the Oppenheimer-Snyder 
 spacetime.}
 \label{fg:rays_os}
\end{figure}

\begin{figure}
 \centerline{\epsfxsize 18cm \epsfysize 14cm \epsfbox{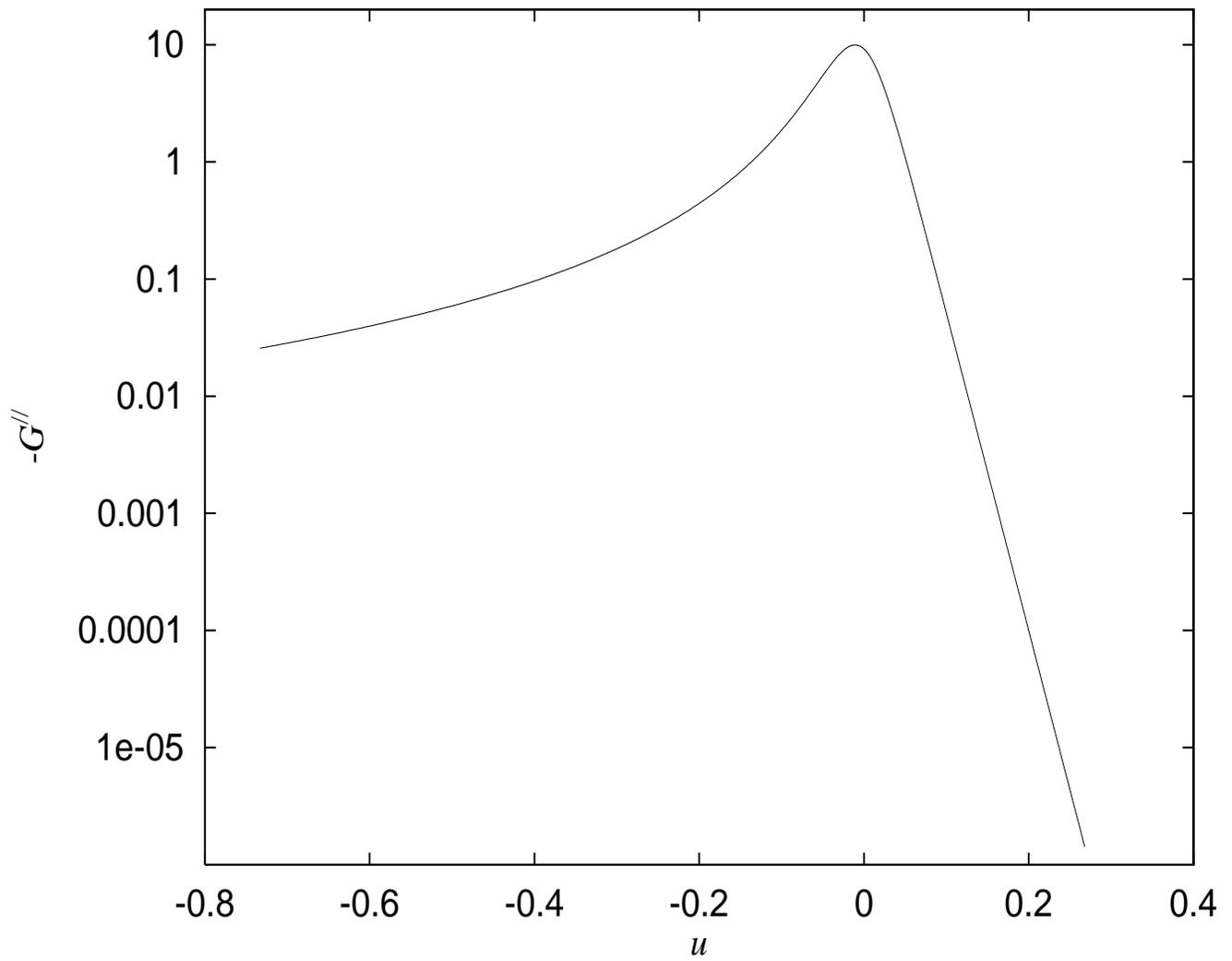}}
 ~(c)
 \caption{Plots of (a) $G(u)$, (b) $G^{\prime}(u)$, and (c)
 $G^{\prime\prime}(u)$ for the Oppenheimer-Snyder spacetime.}
 \label{fg:g_os}
\end{figure}

The final fate of the Oppenheimer-Snyder collapse is a
spacelike singularity covered by the event horizon.
Then it is expected that radiation during the collapse
tends to Hawking radiation as $u\to \infty$.
In reality, this is true.
See Fig.~\ref{fg:rays_os} in which trajectories of null geodesics
are displayed.
We can see that the event horizon covers the singularity.
See Fig.~\ref{fg:g_os} for the behavior of the functions $G(u)$,
$G^{\prime}(u)$, and $G^{\prime\prime}(u)$.
During early times, the behavior of $G(u)$ is basically the same
as that of the globally naked singularity, i.e., $G(u)$ is written as
\begin{equation}
        G(u)\approx u+f_{e}M\ln(-u).
\end{equation}
This is very reasonable because the behavior during early times
will be independent 
of details of the density profile within the dust cloud.
During late times, $G$ is written as follows
\begin{equation}
G(u)\approx -\mbox{const}\times\exp\left(\frac{-u}{4M}\right)+v_{h},
\label{eq:gos}
\end{equation}
where an ingoing null ray with $v_{h}$ 
is reflected to the outgoing 
null ray which is on the event horizon.
This relation will be derived later.

\begin{figure}[htbp]
 \centerline{\epsfxsize 18cm \epsfysize 14cm \epsfbox{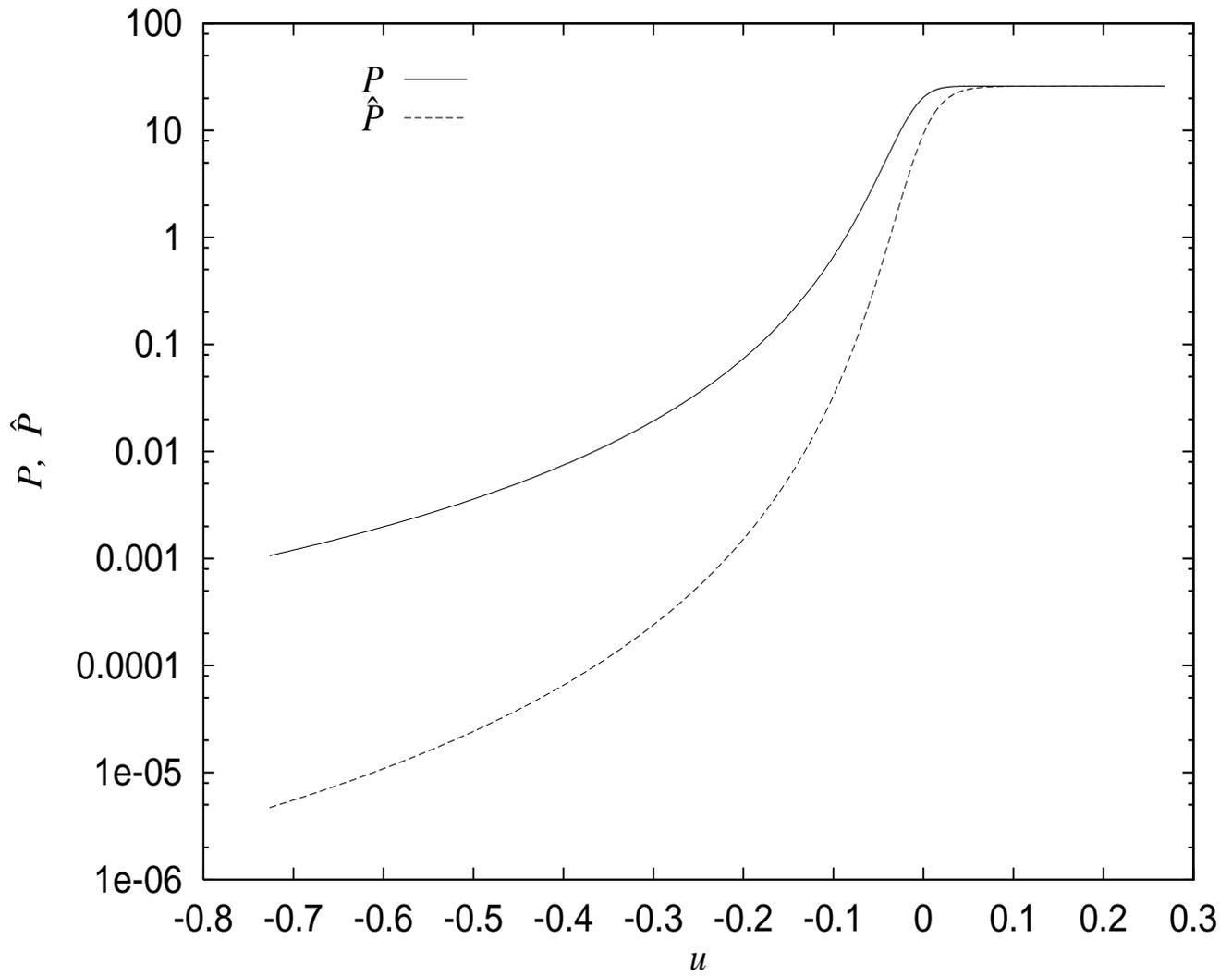}}
 \caption{Power for minimally and conformally
 coupled scalar fields in the Oppenheimer-Snyder spacetime.}
        \label{fg:p_os}
\end{figure}

The power of radiation has been calculated numerically.
During early times, the power is the same 
as that for globally naked singularity.
The results are shown in Fig.~\ref{fg:p_os}.
On the other hand, from Eq.~(\ref{eq:gos}), 
we can estimate the power and 
the radiated energy during late times as
\begin{eqnarray}
P&=&\hat{P}\approx \frac{1}{768\pi M^2}, 
\label{eq:hawkingpower}\\
E&=&\hat{E}\approx \frac{1}{768\pi M^2}u.
\end{eqnarray}
The numerical results show a 
good agreement with these analytic results.  
The numerical result for the power of radiation in this model
has reproduced that of the 
Hawking radiation (\ref{eq:hawkingpower}) 
by a black hole within $\sim 0.01$ \% accuracy.

We should also note that the late time radiation during
the LTB collapse with the locally naked singularity is 
the same as that of the Oppenheimer-Snyder collapse.
This is because 
a distant observer cannot see the locally naked singularity.

\section{spectrum}
\label{sec:spectrum}
\subsection{Total spectrum}
\label{subsec:total_spectrum}
Now that we have obtained the function $F(v)$, we can calculate the 
spectrum of radiation by Eqs.~(\ref{eq:alpha}),
(\ref{eq:beta}), and (\ref{eq:nomega}).
Some analytic expressions for the Bogoliubov coefficients are
obtained in Appendix~\ref{sec:analytic_expressions}.
In order to determine the spectrum numerically,
we introduce the Gaussian window function in the Fourier 
transformation as usual.
That is, in place of Eqs.~(\ref{eq:alpha}) and (\ref{eq:beta}), 
we have calculated the Bogoliubov coefficients
\begin{eqnarray}
\tilde{\alpha}_{\omega^{\prime}\omega}&\equiv&\frac{1}{2\pi}
\sqrt{\frac{\omega^{\prime}}{\omega}}\int^{\infty}_{-\infty}
dve^{i\omega F(v) -i\omega^{\prime}v}
\frac{1}{\sqrt{4\pi}\Sigma}
\exp\left[-\left
(\frac{F(v)-u_{c}}{\Sigma}\right)^2\right], 
\label{eq:tilalpha}\\
\tilde{\beta}_{\omega^{\prime}\omega}&\equiv &
-\frac{1}{2\pi}
\sqrt{\frac{\omega^{\prime}}{\omega}}\int^{\infty}_{-\infty}
dve^{-i\omega F(v) -i\omega^{\prime}v}
\frac{1}{\sqrt{4\pi}\Sigma}
\exp\left[-
\left(\frac{F(v)-u_{c}}{\Sigma}\right)^2\right],
\label{eq:tilbeta}
\end{eqnarray}
where $u_{c}$ and $\Sigma$ are the center 
and the width of the Gaussian window function, respectively.
Since the wave packet
\begin{equation}
   \frac{1}{\sqrt{4\pi}\Sigma}e^{i\omega u}
\exp\left[-\left(\frac{u-u_{c}}{\Sigma}\right)^2\right]
\end{equation}
has the frequency band width $\Delta \omega\sim 2/\Sigma$,
we can obtain the number of particles per unit frequency band as
\begin{equation}
\frac{d N}{d\omega}\sim\frac{\Sigma}{2}
\left[\frac{\int_{0}^{\infty}d\omega^{\prime}
|\tilde{\alpha}_{\omega^{\prime}\omega}|^2}
{\int_{0}^{\infty}d\omega^{\prime}
|\tilde{\beta}_{\omega^{\prime}\omega}|^2}-1\right]^{-1},
\label{eq:finitenumber}
\end{equation}
where we have again 
assumed that the geometrical optics approximation is valid.

Although we could use the calculated data for the function $F(v)$,
we have used the analytic formula which have been derived based on
the numerical results for 
convenience of retaining accuracy.
Since the diverging power
is associated with the late time behavior, 
we concentrate on the late time radiation.
In order to obtain the total spectrum, we extrapolate the 
function $F$ linearly as
\begin{equation}
F(v)=\left\{
\begin{array}{ll}
A^{-1}(v-v_{0})+u_{0} & \qquad (v_{0}<v), \\
A^{-1}(v-v_{0})+\displaystyle{\frac{4}{3}}
f_{l}\omega_{s}^{1/2}[A^{-1}(v_{0}-v)]^{3/2}
+u_{0} &\qquad (v_{1}<v\le v_{0}),
\end{array}\right.
\label{eq:extrapolate_linear}
\end{equation}
where 
$v_{1}$ is given by 
\begin{equation}
v_{0}-v_{1}\equiv\left[\frac{1}{2}(1-A)f_{l}^{-1}\right]
^{2}A\omega_{s}^{-1}.
\end{equation}
We should note that Eq.~(\ref{eq:extrapolate_linear}) can be used
for $u_{0}-u\alt \omega_{tp}^{-1}$.
Later we will show that $\omega_{tp}\agt \omega_{s}$ for 
$r_{sf}\ll l_{0}$.
In fact, it turns out that we only have to pay attention to
the spectrum above $\omega_{tp}$.
We have chosen $u_{c}=u_{0}$ and $\Sigma$ to be
rather smaller than $u_{0}-u_{1}$
with $u_{1}=F(v_{1})$, 
for the contribution dominantly comes from $u\approx u_{0}$.
We should note that there is no radiation for $u>u_{0}$ in this 
extrapolation.
We could adopt another extrapolation, for example,
an extrapolation such that $F(v)-u_{0}$ is antisymmetric
with respect to $v=v_{0}$, i.e.,  
\begin{equation}
F(v)=\left\{
\begin{array}{ll}
A^{-1}(v-v_{0})-\displaystyle{\frac{4}{3}}
f_{l}\omega_{s}^{1/2}[A^{-1}(v-v_{0})]^{3/2}
+u_{0} &\qquad (v_{0}<v <2v_{0}-v_{1}),\\
A^{-1}(v-v_{0})+\displaystyle{\frac{4}{3}}
f_{l}\omega_{s}^{1/2}[A^{-1}(v_{0}-v)]^{3/2}
+u_{0} &\qquad (v_{1}<v\le v_{0}),
\end{array}\right.
\label{eq:extrapolate_antisym}
\end{equation}
This antisymmetric extrapolation turns out to
only double the amplitude of the spectrum 
through the linear extrapolation (\ref{eq:extrapolate_linear}).
Therefore, we only present the spectrum through 
the linear extrapolation.
Because the second derivative of $F$ is diverging as $v\to v_{0}$, 
we can only require that the first derivative of $F$ should be
continuous at $v=v_{0}$.
If we allow discontinuity of the first derivative,
radiation due to this discontinuity dominates the 
radiated energy flux, which is out of concern in this paper. 

\begin{figure}[htbp]
 \centerline{\epsfxsize 18cm \epsfysize 14cm \epsfbox{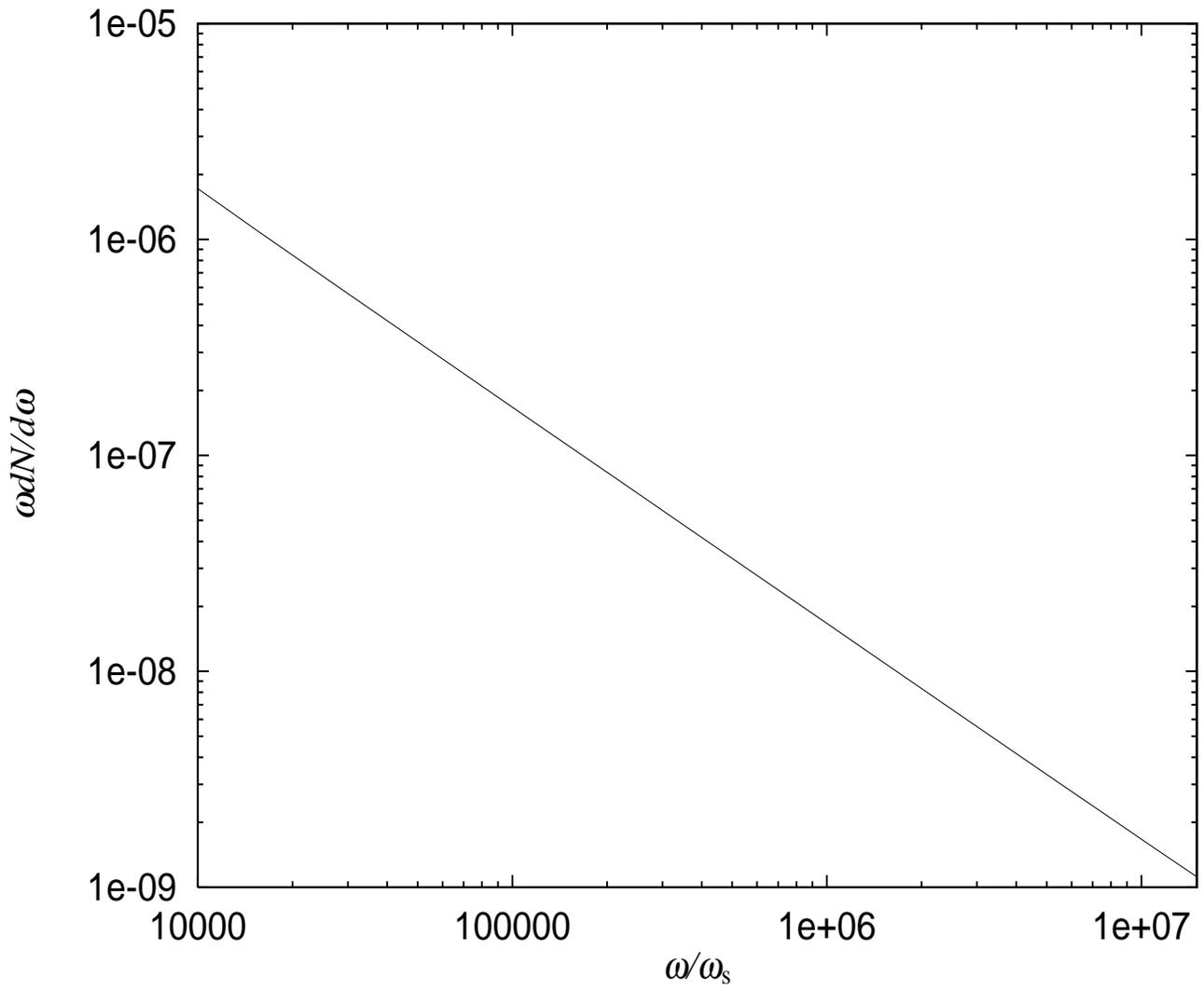}}
 \caption{Total spectrum of a naked singularity explosion for
 $A=0.8$ and $f_{l}=1$.}
 \label{fg:spece_tot}
\end{figure}

The obtained spectrum is shown in Fig.~\ref{fg:spece_tot}.
The parameters are fixed as $A=0.8$ and $f_{l}=1$.
We should note that the contribution to the total 
radiated energy mainly comes from $\omega\agt \omega_{tp}$. 
In this figure, we find 
\begin{equation}
\omega\frac{dN}{d\omega}\propto \omega^{-1}.
\end{equation}
Therefore, the total energy, which will be obtained by
\begin{equation}
E=\int^{\infty}_{0}d\omega \omega \frac{dN}{d\omega},
\end{equation}
is logarithmically divergent, which is 
consistent with the diverging radiated energy obtained
in Sec.~\ref{sec:power} based on
the point-splitting regularization.

For $u>u_{0}$, if we respect the causal structure of 
the background LTB spacetime, in which the naked singularity
is ingoing null, an ingoing null ray never 
changes into an outgoing null ray.
It is obvious that the geometrical 
optics approximation will be no longer 
valid after the formation of the naked singularity.
Therefore, we need to determine mode functions
by other methods than the geometrical optics approximation.  
On the other hand, we have seen that the unbounded power
and energy are radiated before the occurrence of the naked singularity.
Then, we could expect that naked singularity
formation will be prevented for some physical mechanism.
For example, the back reaction of the quantum effect
might prevent 
naked singularity formation.
In fact, the LTB spacetime is only an approximation
of realistic gravitational collapse.
Violation of spherical symmetry, 
rotational support,
hardening of an equation of state,
ignition of nuclear burning,
quantum gravity, and so on
might make the approximation by the LTB spacetime
no longer valid in the final stage of the realistic collapse.
From the above discussions, the causal structure of the LTB spacetime 
after naked singularity formation cannot be taken seriously.

\subsection{Momentary spectrum}
\label{subsec:momentary_spectrum}
It is important in estimating the validity of the geometrical optics
approximation to examine which frequency band dominates the power
at some moment.
For this purpose, we determine the momentary spectrum
by wavelet analysis.
We use Gabor's mother 
wavelet\footnote{The Gabor wavelet is sometimes
referred to as the Morlet wavelet. Strictly speaking, Gabor's mother wavelet 
does not become a basis function. However, it is known that this wavelet
with an appropriate value of $\sigma$ is useful in finding
the frequency of the signal.}
in place of $e^{i x}$
because it has a very clear physical meaning 
as a Gaussian wave packet.
We consider Gabor's mother wavelet 
\begin{equation}
\psi(x)=\frac{1}{\sqrt{4\pi}\sigma}e^{ix}e^{-x^{2}/\sigma^2},
\end{equation} 
where we have set $\sigma=8$.
The result below is not so sensitive to the value of $\sigma$.
We can obtain the Bogoliubov coefficients as
\begin{eqnarray}
\bar{\alpha}_{\omega^{\prime}\omega}&\equiv&\frac{1}{2\pi}
\sqrt{\frac{\omega^{\prime}}{\omega}}\int^{\infty}_{-\infty}
dv\psi[\omega(F(v)-u)]	 e^{-i\omega^{\prime}v},
\label{eq:baralpha}\\
\bar{\beta}_{\omega^{\prime}\omega}&\equiv &
-\frac{1}{2\pi}
\sqrt{\frac{\omega^{\prime}}{\omega}}\int^{\infty}_{-\infty}
dv \psi^{*}[\omega(F(v)-u)] e^{-i\omega^{\prime}v},
\label{eq:barbeta}
\end{eqnarray}
where $u$ is chosen to be the observation time.
Because the Gabor wavelet occupies 
area $\Delta \omega \Delta u \sim 2$
in the phase space $(u,\omega)$, 
the momentary number spectrum is given by
\begin{equation}
\frac{dN}{d\omega du}\sim\frac{1}{2}\Gamma_{\omega}
\left[\frac{\int_{0}^{\infty}d\omega^{\prime}
|\bar{\alpha}_{\omega^{\prime}\omega}|^2}
{\int_{0}^{\infty}d\omega^{\prime}
|\bar{\beta}_{\omega^{\prime}\omega}|^2}-1\right]^{-1},
\end{equation}
where $\Gamma_{\omega}$ is the transmission coefficient,
which we tentatively assume $\Gamma_{\omega}=1$. 

\begin{figure}[htbp]
 \centerline{\epsfxsize 18cm \epsfysize 14cm \epsfbox{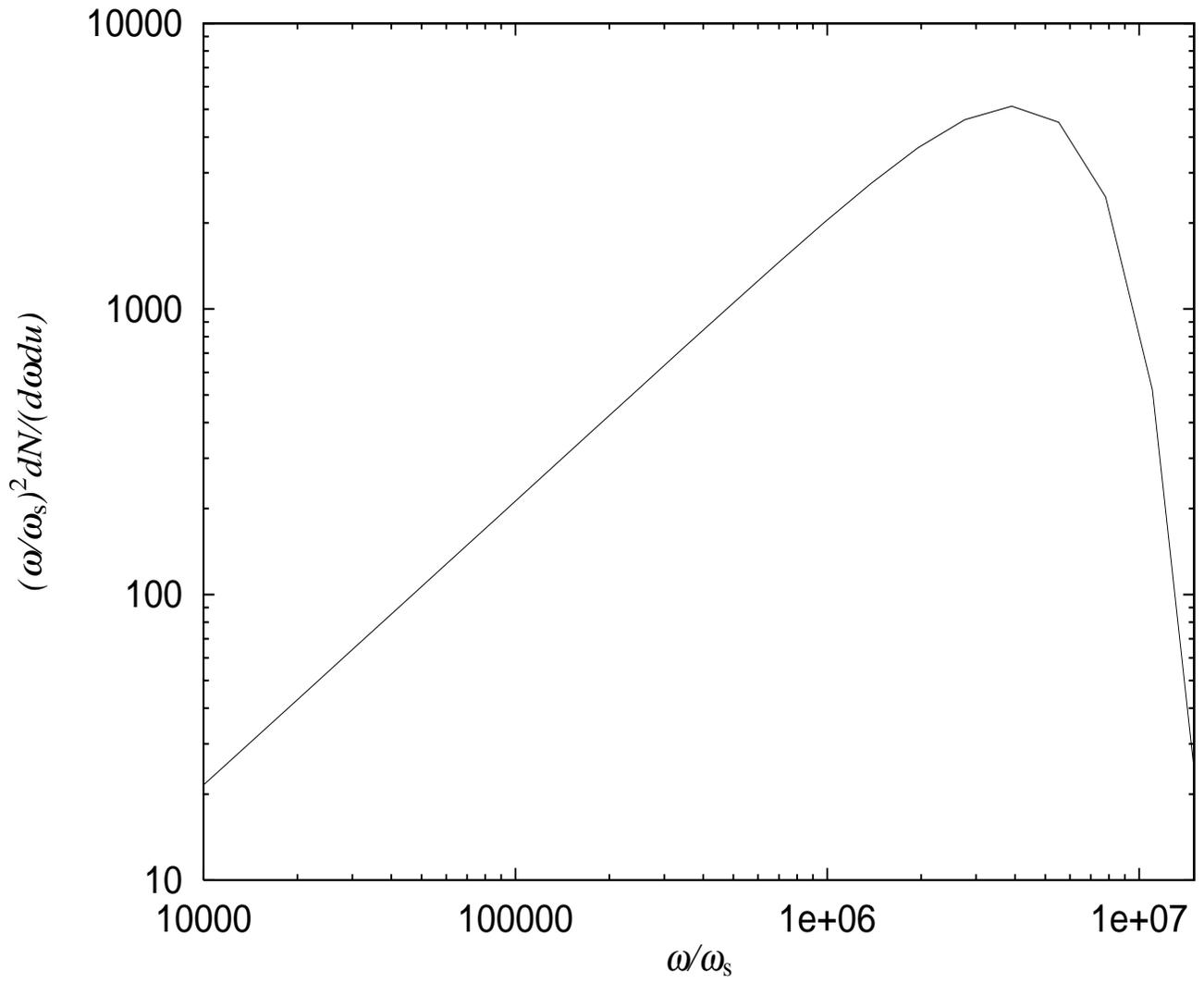}}
 \caption{Momentary spectrum of a naked singularity explosion
at $u_{0}-u=10^{-6}\omega_{s}^{-1}$ for $A=0.8$ and $f_{l}=1$.}
 \label{fg:spece_mom}
\end{figure}

We have calculated numerically the momentary spectrum.
The result is shown in Fig.~\ref{fg:spece_mom}.
This figure displays the contribution of each logarithmic bin  
of frequency to the power at the moment.
The parameters are fixed as $A=0.8$ and $f_{l}=1$.
We have chosen the time of observation $u=10^{-6}\omega_{s}^{-1}$.
The shape of the spectrum looks somewhat 
like the thermal one but
the normalization of the spectrum is different. 
In this figure, it is found that
the contribution to the power
is dominated by the frequency $\omega\sim 2\pi (u_{0}-u)^{-1}$. 
For a conformally coupled scalar field,
the integration of the obtained $\omega dN/(d\omega du)$ with $\omega$
recovers the power derived 
by the point-splitting regularization 
except for a numerical factor
of order unity, i.e.,
\begin{equation}
\int d\omega \omega\frac{dN}{d\omega du} \sim \hat{P}.
\end{equation} 
For a minimally coupled scalar field, such 
a correspondence cannot be seen.
For a conformally coupled scalar field,
since the power is mainly transported 
by particles with a frequency of about $2\pi (u_{0}-u)^{-1}$,
the radiated number of particles per unit time is constant,
being roughly estimated from Eq.~(\ref{eq:hatl}) as
\begin{equation}
\frac{dN}{du}\sim\omega_{s}.
\end{equation} 
From this equation, we can roughly estimate as 
\begin{equation}
\omega\frac{dN}{d\omega}\sim \frac{\omega_{s}}{\omega},
\label{eq:odndo}
\end{equation}
which recovers the total spectrum except for a numerical factor.
From the above analysis, we can conclude that the total divergence 
term in the expression for the power of a minimally coupled
scalar field has nothing to do with the observed particle number.

\section{null geodesics and redshifts in LTB spacetime}
\label{sec:null_geodesics}
\subsection{Assumption}
\label{subsec:assumption}
Here we analytically examine null geodesics and redshifts 
in the LTB spacetime
and derive some useful results to comprehend the problem.
We assume the function $F(r)$ of the form (\ref{eq:expansion})
with $F_{3}>0$ and $F_{5}<0$.
From Eq.~(\ref{eq:R}), $R$ and $R_{,r}$ are written
for $r/l_{0}\ll 1$ as
\begin{eqnarray}
R&\approx& r\left[\left(\frac{-\eta}{t_{0}}\right)
\left\{1-\frac{1}{2}\left(\frac{r}{l_{0}}\right)^2\right\}
+\frac{1}{2}\left(\frac{r}{l_{0}}\right)^2\right]^{2/3}, 
\label{eq:Rexp}\\
R_{,r}&\approx& 
\left[\left(\frac{-\eta}{t_{0}}\right)
\left\{1-\frac{7}{6}\left(\frac{r}{l_{0}}\right)^2\right\}
+\frac{7}{6}\left(\frac{r}{l_{0}}\right)^2\right]
\left[\left(\frac{-\eta}{t_{0}}\right)
\left\{1-\frac{1}{2}\left(\frac{r}{l_{0}}\right)^2\right\}
+\frac{1}{2}\left(\frac{r}{l_{0}}\right)^2\right]^{-1/3},
\label{eq:rpexp}
\end{eqnarray}
where $\eta$ is defined as
\begin{equation}
\eta\equiv t-t_{0}.
\label{eq:eta}
\end{equation}
Moreover, $\eta_{s}(r)\equiv t_{s}(r)-t_{0}$ and 
$\eta_{ah}(r)\equiv t_{ah}(r)-t_{0}$ are 
approximated for $r/l_{0}\ll 1$ as
\begin{eqnarray}
\frac{\eta_{s}(r)}{t_{0}}&\approx &
\frac{1}{2}\left(\frac{r}{l_{0}}\right)^2,
\label{eq:etas}\\
\frac{\eta_{ah}(r)}{t_{0}}&\approx& 
\frac{1}{2}\left(\frac{r}{l_{0}}\right)^2
-\left(\frac{2}{3}\right)^3 
\left(\frac{l_{0}}{t_{0}}\right)^3
\left(\frac{r}{l_{0}}\right)^{3}.
\label{eq:etaah}
\end{eqnarray}

For simplicity, we assume
\begin{equation}
   \frac{r_{sf}}{l_{0}}\ll 1.
\end{equation}
This implies that we can safely expand in powers of $r/l_{0}$
and take the leading-order term {\it in the whole of the cloud}. 

\subsection{Trajectories}
\label{subsec:trajectories}
First we consider the region where 
\begin{equation}
\left(\frac{r}{l_{0}}\right)^2\ll \left|\frac{\eta}{t_{0}}\right|
\label{eq:A}
\end{equation}
is satisfied.
This region is approximately recognized as the Friedmann universe.
From Eq.~(\ref{eq:etas}), it is found that this condition can 
be satisfied only for $\eta<0$.
In this case, $R_{,r}$ is approximated as
\begin{equation}
R_{,r}\approx\left(\frac{-\eta}{t_{0}}
\right)^{2/3}.
\end{equation}
The ordinary differential equation (\ref{eq:dtdr})
can be easily integrated to 
\begin{equation}
\left(\frac{-\eta}{t_{0}}\right)^{1/3}
\approx \mp \frac{1}{3}\left(\frac{l_{0}}
{t_{0}}\right)\left(\frac{r}{l_{0}}
\right)+C_{A\pm},
\label{eq:nullA}
\end{equation}
where $C_{A\pm}$ is a constant of integration.
It is easily found that each null ray can be drawn
by a parallel transport of the curve with $C_{A\pm}=0$ 
along the $r$-axis direction.

Next we consider the region where
\begin{equation}
\left(\frac{r}{l_{0}}\right)^2\gg \left|\frac{\eta}{t_{0}}\right|
\label{eq:B}
\end{equation}
is satisfied.
This region is not at all approximated by the Friedmann universe.
In this case, $R_{,r}$ is approximated as
\begin{equation}
R_{,r}\approx \frac{7}{2^{2/3}3}
\left(\frac{r}{l_{0}}\right)^{4/3}.
\end{equation}
Then Eq.~(\ref{eq:dtdr}) is integrated to 
\begin{equation}
\frac{\eta}{t_{0}} \approx \pm 2^{-2/3}
\left(\frac{l_{0}}{t_{0}}\right)
\left(\frac{r}{l_{0}}
\right)^{7/3}-C_{B\pm},
\label{eq:nullB}
\end{equation}
where $C_{B\pm}$ is a constant of integration.
It is easily found that each null ray can be drawn
by a parallel transport of the curve with $C_{B\pm}=0$ along 
the $\eta$-axis direction.

\begin{figure}[htbp]
 \centerline{\epsfxsize 18cm \epsfysize 14cm \epsfbox{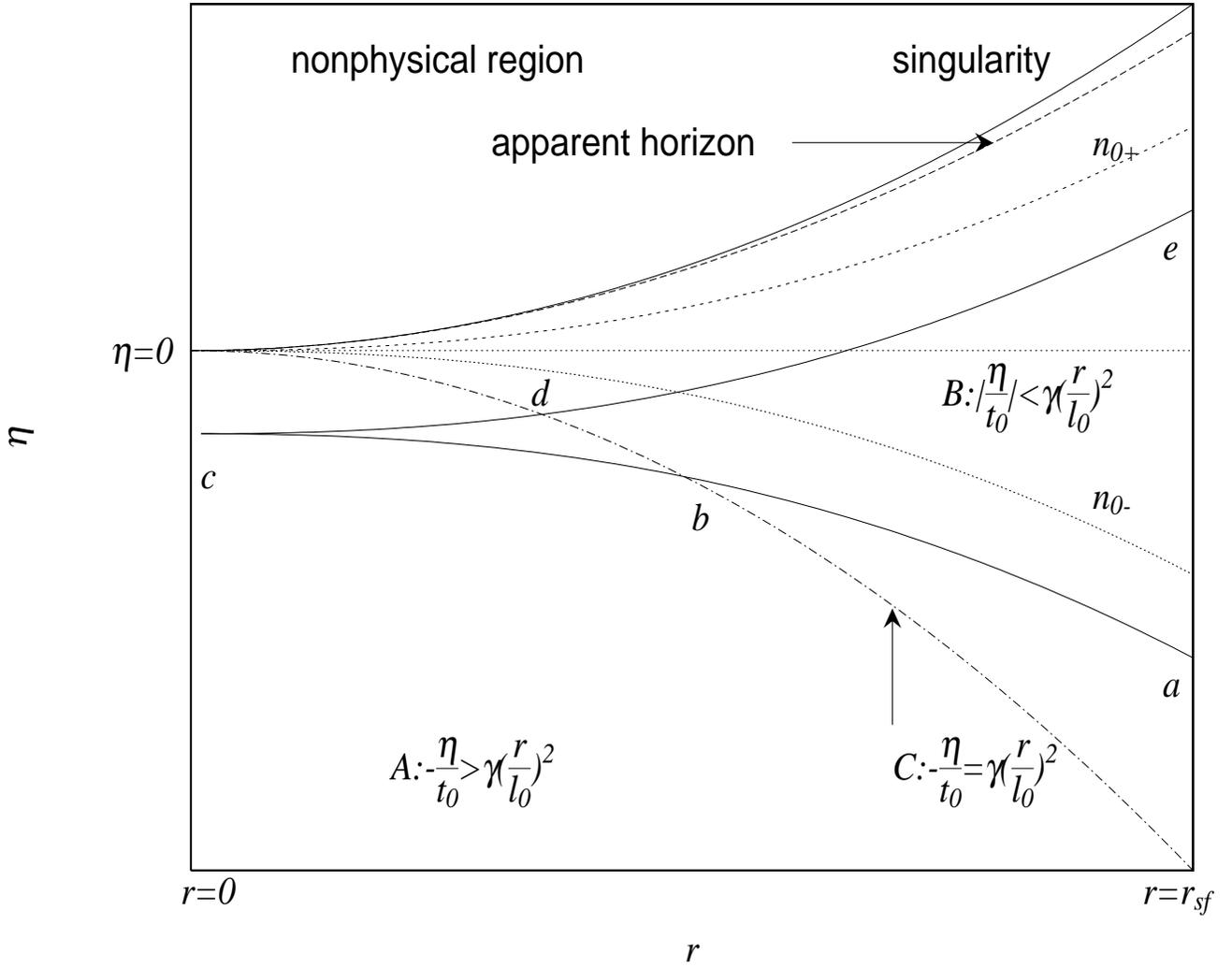}}
 \caption{Schematic diagram of the LTB spacetime interior of the dust 
cloud around the naked singularity}
 \label{fg:sketch}
\end{figure}

See Fig.~\ref{fg:sketch}, which illustrates the spacetime 
around $(\eta,r)=(0,0)$.
Condition (\ref{eq:A}) is satisfied in region $A$, while
condition (\ref{eq:B}) is satisfied in region $B$.
The boundary of regions $A$ and $B$ will be described by
\begin{equation}
\frac{-\eta}{t_{0}}= \gamma \left(\frac{r}{l_{0}}\right)^2,
\label{eq:C}
\end{equation}
where $\gamma$ is a constant of order unity.
We denote this boundary curve as $C$.
It may be kept in mind that this treatment 
is rather simple.
However, we believe that the present approximation,
where we divide the spacetime into two regions,
will be enough to comprehend the essence of the problem.

Then, we concentrate on the behavior of null geodesics
around the naked singularity.
See Fig.~\ref{fg:sketch} for the trajectory of null geodesics.
From the regular center $\eta<0$ and $r=0$, 
the outgoing null ray runs region $A$
which is described by the upper sign of Eq.~(\ref{eq:nullA})
with $C_{A+}>0$.
Then this null ray goes through boundary $C$.
After that the null ray goes into region $B$
and the trajectory is described by the upper sign of 
Eq.~(\ref{eq:nullB}) with  $C_{B+}>0$.
Since all null rays with $C_{B+}>0$ in region $B$
emanate from the regular center, the outgoing null geodesic
with $C_{B+}=0$ in region $B$ generates the Cauchy horizon.
In other words, it is 
the earliest outgoing null ray from the singularity at $r=0$.
We denote this outgoing null geodesic as $n_{0+}$.
There are infinitely many outgoing 
null geodesics later than $n_{0+}$ which emanates from the
naked singularity at $r=0$.
In fact, these null geodesics form one parameter family.
These null rays cannot asymptote to $n_{0+}$ as $r\to 0$ because 
these null rays are obtained by parallel transport 
of $n_{0+}$ in the $\eta$-axis direction in region $B$.
Instead, 
the location of the apparent horizon $\eta=\eta_{ah}(r)$
is an asymptote of these outgoing null rays in approaching
the naked singularity. 

On the other hand,
it is clear that the ingoing null ray with $C_{B-}=0$ 
terminates at the singularity $r=0$.
We denote this ingoing null ray as $n_{0-}$.
For small positive $C_{B-}$, 
the ingoing null ray in region $B$ crosses to boundary $C$
and the null ray becomes described by the
lower sign of Eq.~(\ref{eq:nullA}) with $C_{A-}>0$,
and then terminates at the regular center.
For small negative $C_{B-}$, 
the ingoing null ray in region $B$ 
terminates at the spacelike singularity $r>0$.

\subsection{Condition for globally naked singularity}
\label{subsec:condition}
If $n_{0+}$ reach ${\Im}^{+}$, then the singularity is globally naked,
otherwise the singularity is locally naked.
Noting that the intersection of the apparent horizon 
with the cloud surface 
is on the event horizon, we find that the 
condition for the singularity to 
be globally naked is given by
\begin{equation}
\left[\left(\frac{l_{0}}{t_{0}}
\right)\left(\frac{r_{sf}}{l_{0}}\right)^{1/3}\right]^{-1}
-\frac{16}{27}\left[\left(\frac{l_{0}}{t_{0}}\right)
\left(\frac{r_{sf}}{l_{0}}\right)^{1/3}
\right]^{2}\agt 2^{1/3}.
\end{equation}
Since $x^{-1}-(16/27)x^{2}$ is a decreasing 
function of $x>0$, we can find the condition
for the singularity to be globally naked as
\begin{equation}
\frac{r_{sf}}{l_{0}}\alt 0.66\left(\frac{t_{0}}{l_{0}}\right)^3.
\end{equation}
This implies that the singularity is globally naked for sufficiently
small $r_{sf}$ when $t_{0}$ and $l_{0}$ are fixed.
Thus, it turns out that 
the present assumption $r_{sf}/ l_{0}\ll 1$ is relevant 
for the globally naked singularity case.
We can translate the above condition to a condition 
for $M$ and $\omega_{s}$ as
\begin{equation}
M\alt 6.4\times 10^{-2} \omega_{s}^{-1}.
\label{eq:globallynaked}
\end{equation}
It implies that the mass of the dust cloud with the globally 
naked singularity is bounded from above.

\subsection{Redshifts}
\label{subsec:redshifts}
Since it is necessary for later discussions, 
we turn our attention to redshifts of particles.
Let $k^{\mu}$ be the tangent vector of a radial null geodesic.
Inside of the dust cloud, $k^{\mu}$ satisfies
the following equations:
\begin{eqnarray}
-(k^{t})^2+(R_{,r})^2 (k^{r})^2&=&0, 
\label{eq:null} \\
\frac{dk^{t}}{d\lambda}+R_{,r}{R}_{,tr}(k^{r})^2&=&0, 
\label{eq:geoeta} \\
\frac{d}{d\lambda}\left[(R_{,r})^{ 2}k^{r}\right]
-R_{,r}R_{,rr}(k^{r})^2&=&0,
\label{eq:geor}
\end{eqnarray}
where $\lambda$ is the affine parameter.
The frequency $\hat{\omega}$ observed by an observer comoving 
with a fluid element is
calculated from $k^{\mu}$ as
\begin{equation}
\hat{\omega}=-k^{\mu}u_{\mu}=-k^{t}.
\end{equation}

In region $A$, Eqs.~(\ref{eq:null})-(\ref{eq:geor})
can be integrated as
\begin{eqnarray}
k^{t}(-\eta)^{2/3}&\approx& \mbox{const}, \\
k^{r}(-\eta)^{4/3}&\approx&\mbox{const}
\end{eqnarray}
in the lowest order.
Therefore, a particle is blueshifted for the comoving observer
in region $A$.
In region $B$, Eqs.~(\ref{eq:null})-(\ref{eq:geor})
can be integrated as
\begin{eqnarray}
k^{t}&\approx&\mbox{const}, \\
k^{r}r^{4/3}&\approx&\mbox{const}
\end{eqnarray}
in the lowest order.
Therefore a particle is neither redshifted nor blueshifted
in region $B$.

In the external Schwarzschild spacetime,
in a similar way, we obtain the tangent vector of
a radial null geodesic as
\begin{eqnarray}
k^{T}\left(1-\frac{2M}{R}\right)&=&\mbox{const}, \\
k^{R}&=&\mbox{const}.
\end{eqnarray}
A static distant observer observes the frequency
\begin{equation}
\omega=-k^{T}(T,\infty).
\end{equation}
Using the matching conditions (\ref{eq:match1})-(\ref{eq:match3}),
the observed frequency $\hat{\omega}=-k^{\mu}u_{\mu}$ by 
the comoving observer
at the surface is written using the above $\omega$ as
\begin{equation}
\hat{\omega}=\frac{\omega}{1\mp\alpha},
\label{eq:redshiftfactor}
\end{equation}
where $\alpha$ is defined by
\begin{equation}
\alpha\equiv\sqrt{\frac{2M}{R}},
\end{equation}
and $R$ is the areal radius of the dust surface 
when the null ray crosses the surface.

\subsection{Estimate of function $G$ or $F$}
\label{subsec:estimate}
Let $\alpha_{0\pm}$ be the value of $\alpha$ for $n_{0\pm}$.
Let $\eta_{0\pm}$ be the value of $\eta$ 
when $n_{0\pm}$ crosses the dust surface.
Then, from Eq.~(\ref{eq:v}), 
it is easy to see that the Taylor expansion is valid
around $\eta=\eta_{0-}$ for $v$ as a function of $\eta$ as
\begin{equation}
v=v_{0}+\frac{1}
{1+\alpha_{0-}}(\eta-\eta_{0-})+O\left((\eta-\eta_{0-})^2\right).
\label{eq:taylorv}
\end{equation}
For $\alpha_{0+}<1$, from Eq.~(\ref{eq:u}), it is also 
easy to see that the Taylor expansion is valid
around $\eta=\eta_{0+}$ for $u$ as a function of $\eta$ as
\begin{equation}
u=u_{0}+\frac{1}
{1-\alpha_{0+}}(\eta-\eta_{0+})+O\left((\eta-\eta_{0+})^2\right),
\label{eq:tayloru}
\end{equation}
or equivalently,
\begin{equation}
\eta_{0+}-\eta=(1-\alpha_{0+})
(u_{0}-u)+O((u_{0}-u)^2).
\label{eq:tayloreta}
\end{equation}

See Fig.~\ref{fg:sketch} again.
An ingoing light ray which is close to $n_{0\pm}$ enters the dust 
cloud in region $B$ ($a$),
enters region $A$ ($b$), 
crosses the center ($c$), becomes an outgoing light ray,
enters again region $B$ ($d$) and
leaves the dust cloud ($e$). 
In order to determine $G$ or $F$, we 
need to relate the time coordinate
$\eta_{a}$ at which the incoming 
light ray enters the dust cloud with
$\eta_{e}$ at which the outgoing 
light ray leaves the dust cloud.
Within the present approximation, the following relations 
hold:
\begin{eqnarray}
\frac{\eta_{0-}-\eta_{a}}{t_{0}}&\approx&
\frac{-\eta_{b}}{t_{0}}-2^{-2/3}\gamma^{-7/6}
\left(\frac{l_{0}}{t_{0}}\right)
\left(\frac{-\eta_{b}}{t_{0}}\right)^{7/6}, \\
\left(\frac{-\eta_{c}}{t_{0}}\right)^{1/3}&\approx&
\left(\frac{-\eta_{b}}{t_{0}}\right)^{1/3}
-\frac{1}{3}\gamma^{-1/2}\left(\frac{l_{0}}{t_{0}}\right)
\left(\frac{-\eta_{b}}{t_{0}}\right)^{1/2}, \\
\left(\frac{-\eta_{c}}{t_{0}}\right)^{1/3}&\approx&
\left(\frac{-\eta_{d}}{t_{0}}\right)^{1/3}
+\frac{1}{3}\gamma^{-1/2}\left(\frac{l_{0}}{t_{0}}\right)
\left(\frac{-\eta_{d}}{t_{0}}\right)^{1/2}, \\
\frac{\eta_{0+}-\eta_{e}}{t_{0}}&\approx&
\frac{-\eta_{d}}{t_{0}}+2^{-2/3}\gamma^{-7/6}
\left(\frac{l_{0}}{t_{0}}\right)
\left(\frac{-\eta_{d}}{t_{0}}\right)^{7/6},
\end{eqnarray}
where $\eta_{0\pm}$ are written as
\begin{equation}
\eta_{0\pm}\approx\pm 2^{-2/3}\left(\frac{l_{0}}{t_{0}}\right)
\left(\frac{r_{sf}}{l_{0}}\right)^{7/3}.
\end{equation}
From the above relations, we find
\begin{equation}
\eta_{0+}-\eta_{e}\approx \eta_{0-}-\eta_{a}.
\end{equation}

Then, we find for $\alpha_{0+}<1$
\begin{equation}
v=G(u)\approx v_{0}-\frac{1-\alpha_{0+}}{1+\alpha_{0-}}
(u_{0}-u),
\end{equation}
or
\begin{equation}
u=F(v)\approx u_{0}-\frac{1+\alpha_{0-}}{1-\alpha_{0+}}(v_{0}-v).
\end{equation}
From the above discussions, we can determine 
$A=\lim_{u\to u_{0}}G^{\prime}$ as
\begin{equation}
A=\frac{1-\alpha_{0+}}{1+\alpha_{0-}}.
\end{equation}

For the collapse to a black hole,
from Eq.~(\ref{eq:u}), we can express $u$ by $\eta$ around $R=2m$ 
at the surface as 
\begin{equation}
u\approx -4M\ln(\eta_{h}-\eta)+\mbox{const},
\end{equation}
where $\eta_{h}$ is the time when the event horizon crosses 
the dust surface.
On the other hand, we can easily find the following linear relation
when an incoming light ray with the 
advanced time $v$ becomes to an outgoing
null ray which reaches the surface at the time $\eta$ as
\begin{equation}
        \eta_{h}-\eta\approx \mbox{const}\times(v_{h}-v),
\end{equation}
where the ingoing null ray with $v_{h}$ is
reflected to the outgoing ray on the event horizon.
Then we find Eq.~(\ref{eq:gos}).

\section{validity of the geometrical optics approximation}
\label{sec:validity}
The geometrical optics approximation is exact for 
the two-dimensional spacetime,
where the line element is given by setting $d\Omega^{2}=0$
in the four-dimensional line element (\ref{eq:metric}).
Therefore, the power, energy, and spectrum obtained
in Secs.~\ref{sec:power} and~\ref{sec:spectrum} 
are exact in this sense for the two-dimensional version of 
the LTB spacetime.

In the four-dimensional case, the geometrical 
optics approximation is only an
approximation because of the existence of the curvature potential.
The geometrical optics approximation is valid for waves 
with the wavelength
shorter than the curvature radius of the spacetime geometry.
In other words, the geometrical optics approximation is good if 
condition \begin{equation}
\hat{\omega}\agt 2\pi 
|R_{\hat{\alpha}\hat{\beta}\hat{\gamma}\hat{\delta}}|^{1/2}
\end{equation}
is satisfied,
where the hat denotes components 
in the local inertial tetrad frame.
The nonvanishing components of the Riemann tensor are as follows:
\begin{eqnarray}
R_{\hat{t}\hat{r}\hat{t}\hat{r}}&=&
-R_{\hat{t}\hat{r}\hat{r}\hat{t}}=
-R_{\hat{r}\hat{t}\hat{t}\hat{r}}=
R_{\hat{r}\hat{t}\hat{r}\hat{t}}=-\frac{{R}_{,ttr}}
{R_{,r}}, \\
R_{\hat{t}\hat{\theta}\hat{t}\hat{\theta}}&=&
-R_{\hat{t}\hat{\theta}\hat{\theta}\hat{t}}=
-R_{\hat{\theta}\hat{t}\hat{t}\hat{\theta}}=
R_{\hat{\theta}\hat{t}\hat{\theta}\hat{t}}
=R_{\hat{t}\hat{\phi}\hat{t}\hat{\phi}}=
-R_{\hat{t}\hat{\phi}\hat{\phi}\hat{t}}=
-R_{\hat{\phi}\hat{t}\hat{t}\hat{\phi}}=
R_{\hat{\phi}\hat{t}\hat{\phi}\hat{t}}=
-\frac{R_{,tt}}{R}, \\
R_{\hat{r}\hat{\theta}\hat{r}\hat{\theta}}&=&
-R_{\hat{r}\hat{\theta}\hat{\theta}\hat{r}}=
-R_{\hat{\theta}\hat{r}\hat{r}\hat{\theta}}=
R_{\hat{\theta}\hat{r}\hat{\theta}\hat{r}}
=R_{\hat{r}\hat{\phi}\hat{r}\hat{\phi}}=
-R_{\hat{r}\hat{\phi}\hat{\phi}\hat{r}}=
-R_{\hat{\phi}\hat{r}\hat{r}\hat{\phi}}=
R_{\hat{\phi}\hat{r}\hat{\phi}\hat{r}}=
\frac{R_{,t}R_{,tr}}
{RR_{,r}}, \\
R_{\hat{\theta}\hat{\phi}\hat{\theta}\hat{\phi}}&=&
-R_{\hat{\theta}\hat{\phi}\hat{\phi}\hat{\theta}}=
-R_{\hat{\phi}\hat{\theta}\hat{\theta}\hat{\phi}}=
R_{\hat{\phi}\hat{\theta}\hat{\phi}\hat{\theta}}=
\left(\frac{R_{,t}}{R}\right)^2.
\end{eqnarray}

In region $A$, they are approximated as
\begin{equation}
|R_{\hat{t}\hat{r}\hat{t}\hat{r}}
|\approx|R_{\hat{t}\hat{\theta}\hat{t}\hat{\theta}}|
\approx|R_{\hat{r}\hat{\theta}\hat{r}\hat{\theta}}
|\approx|R_{\hat{\theta}\hat{\phi}\hat{\theta}\hat{\phi}}|
\approx(-\eta)^{-2}.
\end{equation}
In region $B$, they are approximated as
\begin{equation}
|R_{\hat{t}\hat{r}\hat{t}\hat{r}}|\approx
|R_{\hat{t}\hat{\theta}\hat{t}\hat{\theta}}|
\approx|R_{\hat{r}\hat{\theta}\hat{r}\hat{\theta}}|
\approx|R_{\hat{\theta}\hat{\phi}\hat{\theta}\hat{\phi}}|\approx
\frac{1}{t_{0}^2}\left(\frac{r}{l_{0}}\right)^{-4}.
\end{equation}
In Sec.~\ref{subsec:redshifts}, we have seen that
$\hat{\omega}$ is kept constant approximately in region $B$
and that
$\hat{\omega}$ may be considerably blueshifted in region $A$.
However, in Sec.~\ref{subsec:estimate}, we can see
that the null ray which goes into the dust cloud
enters region $A$ and get out of region $A$
at the same time in the lowest order, i.e,
$\eta_{b}\approx \eta_{c}\approx \eta_{d}$.
This implies that the frequency of such a particle is kept
almost constant all over the dust cloud.
On the other hand, the Riemann tensor along the 
light ray reaches the maximum 
when the light ray goes through region $A$. 
Now we can write down the condition for the geometric
optics approximation to be valid as
\begin{equation}
\hat{\omega}\agt 2\pi(-\eta_{c})^{-1}.
\end{equation}
This condition can be rewritten by the quantities on ${\Im}^{+}$
using Eqs.~(\ref{eq:redshiftfactor}) and (\ref{eq:tayloru}).
The result is
\begin{equation}
\omega\agt \omega_{cr}(u)\equiv 2\pi (u_{0}-u)^{-1}
\end{equation}
for $\alpha_{0+}<1$.

As we have shown in Sec.~\ref{sec:spectrum},
the power for each mode is mainly transported 
by particles of which the frequency is about $2\pi (u_{0}-u)^{-1}$.
Since the transmission coefficient $\Gamma_{\omega}$
will be a function of $(\omega/\omega_{cr})$,
it is natural to estimate $\Gamma_{\omega}(\omega=\omega_{cr})$ 
as of the order unity.
It implies that the calculations based on
the geometrical optics approximation will be 
valid for a rough order estimate.  

\section{discussions}
\label{sec:discussions}
Here we compare the naked singularity explosion with the famous 
radiative phenomena, the Hawking radiation, and 
the black hole evaporation.
The Hawking radiation is derived in the fixed background spacetime
as has been done in this paper.
The result is thermal radiation, the temperature of which is given by
\begin{equation}
        T=\frac{1}{8\pi M},
\end{equation}
and the power is given by
\begin{equation}
        P=\hat{P}=\frac{1}{768\pi M^2}.
\end{equation}
The Hawking radiation is not explosive but 
constant.

In contrast to the naked singularity explosion,
it is not until the back reaction of the 
quantum effect is taken into
account that explosive radiation is emitted. 
For the black hole evaporation, the power is given by
\begin{equation}
P\sim (t_{ev}-t)^{-2/3},
\label{eq:evaporationp} 
\end{equation}
where $t_{ev}$ is the time of evaporation and the above
equation is valid only for $t_{ev}-t>1$, where $1$ means, of course,
the Planck time.
The divergence is apparently much milder
than the naked singularity explosion.
The expression of the power does not have
any parameter dependence.
The radiated energy during the final evaporation phase is
only finite in contrast to the naked singularity. 
The mass of the evaporating black hole is given by
\begin{equation}
M\sim (t_{ev}-t)^{1/3}.
\end{equation}
From the above equations, the power at the time $t$ is dominated 
by particles with frequency 
\begin{equation}
\omega \sim (t_{ev}-t)^{-1/3}.
\label{eq:evaporationo}
\end{equation}
The total spectrum is estimated 
from Eqs. (\ref{eq:evaporationp}) and 
(\ref{eq:evaporationo}) as
\begin{equation}
\omega\frac{dN}{d\omega}\sim \omega^{-2},
\label{eq:evaporations}
\end{equation}
where this estimate is valid for $\omega<1$.
Equation~(\ref{eq:evaporations}) implies that 
the spectrum is apparently softer than 
that of the naked singularity explosion.
For the Hawking radiation, since the Schwarzschild
black hole has only one parameter,
i.e., its mass, the back reaction effect is 
rather easy to be taken into account 
although it is phenomenological.
For the naked singularity, the system is dynamical
and not parametrized by only one parameter.
It suggests that it is not an easy task to take the 
back reaction of the quantum effect into account
even phenomenologically.
The comparison of radiation processes in the collapse of a dust ball
is shown in Table \ref{tb:comparison}. 

\begin{table}[htbp]
\begin{center}
  \caption{Comarison of radiation processes: Hawking radiation (HR), black hole
evaporation (BHE), naked singularity explosion 
(NSE)~\protect\cite{hin2000}, and naked singularity 
explosion in self-similar spacetime 
(NSESS)~\protect\cite{bsvw1998a,bsvw1998b,vw1998}.}
  \label{tb:comparison}
    \begin{tabular}{l|cccc} 
      Process  & HR & BHE & NSE & NSESS
      \\ \hline
      Radiation   & Constant & Explosive & Explosive & Explosive \\
      $P$ (total div. subtracted) & $(768\pi M^2)^{-1}$ & $\sim (t_{ev}-t)^{-2/3}$ 
      & $\sim \nu_{s}(t_{\rm CH}-t)^{-1}$ & $\sim (t_{\rm CH}-t)^{-2}$ \\
      $E$ & $\sim (768\pi M^2)^{-1} t$ & $M$ 
      & $\sim \nu_{s}\log (t_{\rm CH}-t)$ & $\sim (t_{\rm CH}-t)^{-1}$ \\ 
      $\nu \displaystyle{\frac{dN}{d\nu}}$ 
      & $\displaystyle{\frac{\nu}{e^{16\pi^{2} M\nu}-1}}$ 
      & $\sim \nu^{-2}$ & $\sim \left(\displaystyle{\frac{\nu}
        {\nu_{s}}}\right)^{-1}$ & ? \\
      $<\nu>$ & $\sim M^{-1}$ & $\sim (t_{ev}-t)^{-1/3}$ & 
        $\sim (t_{\rm CH}-t)^{-1}$ & ? \\ 
      Back reaction &   & $\displaystyle{\frac{dM}{dt}}=-P$  &   &   \\
      Geometrical approx. & Good & Good & Marginal & ? \\     
      Redshift & $\infty$ & $\infty$ & Finite & $\infty$ 
    \end{tabular}
  \end{center}
\end{table}

Again we return to the naked singularity explosion.
The singularity frequency $\omega_{s}$ is obtained from 
Eq.~(\ref{eq:omegas}) as
\begin{eqnarray}
\omega_{s}\sim
10^{-3}
\left(\frac{t_{0}}{1\mbox{ms}}\right)^{-7}
\left(\frac{l_{0}}{10\mbox{km}}\right)^{6} ~\mbox{Hz}.
\end{eqnarray}
The characteristic frequency depends not on   ly on the free fall time of
the system but also greatly on the scale of inhomogeneity.
The scale of inhomogeneity becomes larger, 
the singularity frequency becomes higher.
In some situations, gravitational collapse will set in
due to the growth of small
perturbations of hydrostatic equilibrium in a Newtonian regime.
In such a case, $t_{0}$ and $l_{0}$ will be related as
$l_{0}/t_{0}=\sqrt{15}c_{s}$,
where $c_{s}$ is the sound speed at the center.
Then the singularity frequency is given by
$\omega_{s}\sim 10^{3}  c_{s}^{6} t_{0}^{-1}$.
It implies that $\omega_{s}$ does not much exceed the inverse of 
the initial free fall
time in such situations. 

From Eq.~(\ref{eq:odndo}), it is found that
\begin{equation}
N\sim\int^{\infty}_{\omega_{tp}}d\omega \frac{dN}{d\omega}
\sim \frac{\omega_{s}}{\omega_{tp}}\sim (M\omega_{s})^{2/3} \alt 1.
\end{equation}
Thus, the total number of emitted particles with 
frequency above $\omega_{tp}$ is a few at most.
It suggests that the radiation may be extremely directional
because a few particles transport almost
all the energy radiated during the collapse. 

We are not sure how to take into account the back reaction of 
the quantum effect.
However, we can assume that the back reaction of the quantum effect
becomes important if a considerable fraction of the 
total energy of the
system are radiated away by the quantum particle creation.
From this assumption, we can estimate the emitted energy as
$E\sim 10^{54}(M/M_{\odot})\mbox{erg}$, although $M$ must satisfy
condition (\ref{eq:globallynaked}).
For $E=f M$ with $f\alt 1$,
the averaged energy $\omega_{av}$ of the 
radiated particles above $\omega_{tp}$ is given by
\begin{equation}
\omega_{av}\sim E/N \sim E(M\omega_{s})^{-2/3} \sim
f M^{1/3}\omega_{s}^{-2/3} .
\end{equation}
The central energy density $\epsilon_{c}$ at the final stage is given by
\begin{equation}
\epsilon_{c}\sim \omega_{tp}^{2}\exp
\left(\frac{2E}{\omega_{s}}\right)
\sim M^{-4/3}\omega_{s}^{2/3}
\exp\left(\frac{2E}{\omega_{s}}\right),
\end{equation}
because $\Delta t$ is given by
\begin{equation}
\Delta t\sim \omega_{tp}^{-1}\exp
\left(-\frac{E}{\omega_{s}}\right)\sim
M^{2/3}\omega_{s}^{-1/2}\exp
\left(-\frac{E}{\omega_{s}}\right).
\end{equation}
Actually, this estimate for the radiated energy should
be considered as an upper bound.
It is noted that we have to transform these 
intrinsic quantities to observed quantities
by considering possible reactions of created energetic particles.

The naked singularity explosion may be a new
candidate for a source of ultrahigh energy cosmic rays
because it has a potentially extremely hard spectrum.
It may also provide a candidate for 
a central engine of $\gamma$-ray bursts.
The naked singularity explosion could contribute $\gamma$-ray 
background radiation or disturb big bang nucleosynthesis.

\section{conclusions}
\label{sec:conclusions}
We have considered the naked singularity which occurs in
the collapse of a dust ball with the initial density profile
which is physically reasonable.
We have calculated radiation during the naked singularity formation 
under the geometrical optics approximation.
Since the regularization in these calculations has been done only
in a flat spacetime, the results contain no ambiguity 
which could come from local curvature. 
We have obtained the power and energy
of the radiation, which turn out to diverge to
positive infinity for both minimally and conformally coupled
massless scalar fields.
We have also obtained the spectrum of radiation,
both total and momentary ones.
Then we have found that 
the power is dominantly contributed from
particles of frequency $\omega\sim 2\pi (t_{\rm CH}-t)^{-1}$, where
$(t_{\rm CH}-t)$ is the remaining time until a distant observer
could see the singularity.
The spectrum is given by 
$\omega dN/d\omega\approx (\omega/\omega_{s})^{-1}$,
which is much harder than that of the black hole evaporation.
The radiation depends on the characteristic frequency 
$\omega_{s}$ of the singularity,
which is determined by 
the initial free fall time and the initial 
scale of inhomogeneity at the center.
The radiation may be expected to be strongly beamed.
The geometrical optics approximation has turned out to be consistent
for a rough order estimate.
It will provide a very interesting candidate
for a central engine of $\gamma$-ray bursts or a source of 
ultrahigh energy cosmic rays
because the predicted intrinsic spectrum 
may be extended to an extremely 
high frequency.
A detailed study on such astrophysical 
implications will be a future work.

After this manuscript was received, the paper~\cite{jdm2000}
appeared, in which the naked singularity formation is discussed
as a possible origin of $\gamma$-ray bursts.  
\acknowledgments
We are grateful to H.~Sato for his continuous encouragement.
We are also grateful to T.~Nakamura, H.~Kodama, T.P.~Singh, 
A.~Ishibashi, and S.S.~Deshingkar for helpful discussions.
This work was supported by the 
Grant-in-Aid for Scientific Research (No. 05540)
and for Creative Basic Research (No. 09NP0801)
from the Japanese Ministry of
Education, Science, Sports, and Culture.

\appendix
\section{conserved quantity in LTB collapse}
\label{sec:conserved_quantity}
In Sec.~\ref{sec:LTB}, we set the initial data at $t=0$
and specified the solution.
In fact, the choice of an initial slice is arbitrary.
Therefore, another slice among the
synchronous comoving slicings can be chosen as an initial slice.
Since this variety of choices of an initial slice,
of course, does not change the original collapse model,
physical quantities, such as,
power and energy, must not depend
on the choice of an initial slice.

At an arbitrary time $t<t_{0}$, 
using Eqs.~(\ref{eq:expansion}) and
(\ref{eq:Rexp}),
we can expand the mass function
$F(r)$ in terms of the areal radius $R$ as
\begin{equation}
F(r)=F_{3}\left(\frac{-\eta}{t_{0}}\right)^{-2}R^3
+F_{5}\left(\frac{-\eta}{t_{0}}\right)^{-13/3}R^5+\cdots.
\end{equation} 
We can define the coefficients of $R^3$ and $R^5$ as
$\tilde{F}_{3}(\eta)$ and $\tilde{F}_{5}(\eta)$.
Then we can define the free fall time $\tilde{t}_{0}(\eta)$ 
and the scale of inhomogeneity $\tilde{l}_{0}(\eta)$
at the time $\eta$ as
\begin{eqnarray}
\tilde{t}_{0}(\eta)&\equiv& \frac{2}{3}(\tilde{F}_{3}(\eta))^{-1/2}
=t_{0} \left(\frac{-\eta}{t_{0}}\right), \\
\tilde{l}_{0}(\eta)&\equiv& \left(\frac{-\tilde{F}_{5}(\eta)}
{\tilde{F}_{3}(\eta)}\right)^{-1/2}
=l_{0} \left(\frac{-\eta}{t_{0}}\right)^{7/6}, 
\end{eqnarray}
where $t_{0}$ and $l_{0}$ are understood as $\tilde{t}_{0}(0)$ 
and $\tilde{l}_{0}(0)$,
respectively.
Then we can easily find that
the following quantity is independent of the choice of an 
initial slice.
\begin{equation}
\frac{\tilde{t}_{0}(\eta)^{7}}{\tilde{l}_{0}(\eta)^{6}}
=\frac{t_{0}^{7}}{l_{0}^{6}}
=\mbox{const}.
\end{equation}

\section{analytic expressions of Bogoliubov coefficients}
\label{sec:analytic_expressions}
As seen in Sec.~\ref{sec:null_geodesics}, for $\alpha_{0+}<1$, 
we can write the function $F(v)$ 
in the following form
\begin{equation}
F(v)\approx u_{0}-a(v_{0}-v)+b(v_{0}-v)^{\delta},
\label{eq:Fexpansion}
\end{equation}
where $\delta$, $a$ and $b$ satisfy
\begin{eqnarray}
\delta&=&\frac{3}{2}, \\
a&>&0,\\
b&>&0.
\end{eqnarray}
We denote the advanced time as $v=v_{0}-d$ after which 
the higher-order terms can be neglected,
where $d=O(b^{-1/(\delta-1)})$ is satisfied.
Because the particle creation will occur due to the
third term in the right-hand side of Eq.~(\ref{eq:Fexpansion}),
the dominant contribution to the spectrum comes from 
the integral on $v_{0}-d<v<v_{0}$ in Eqs.~(\ref{eq:alpha})
and (\ref{eq:beta}).
Then, the Bogoliubov coefficients are obtained from 
Eqs.~(\ref{eq:alpha}) and (\ref{eq:beta}) as
\begin{eqnarray}
\alpha_{\omega^{\prime}\omega}&\approx&
\frac{1}{2\pi}\sqrt{\frac{\omega^{\prime}}{\omega}}
e^{i\omega u_{0}-i\omega^{\prime}v_{0}}
\int^{d}_{0}e^{-i\omega(as-b s^{\delta})+i\omega^{\prime}s}ds,\\
\label{eq:alphanaked}
\beta_{\omega^{\prime}\omega}&\approx&
-\frac{1}{2\pi}\sqrt{\frac{\omega^{\prime}}{\omega}}
e^{-i\omega u_{0}-i\omega^{\prime}v_{0}}
\int^{d}_{0}e^{i\omega (as-bs^{\delta})+i\omega^{\prime}s}ds.
\label{eq:betanaked}
\end{eqnarray}
They have the following expression
\begin{eqnarray}
\alpha_{\omega^{\prime}\omega}&\approx&
\frac{1}{2\pi\sqrt{\omega\omega^{\prime}}}
e^{i\omega u_{0}-i\omega^{\prime}v_{0}}
\sum^{\infty}_{k=0}\sum^{k}_{j=0}(-1)^{j}i^{k}
\frac{C^{k}_{~j}}{k!}a^{j}\frac{\omega^kc^{(\delta-1)(k-j)}}
{(\omega^{\prime})^{k+(\delta-1)(k-j)}}
\int^{\omega^{\prime}d}_{0}dx e^{ix}x^{j+\delta(k-j)},
\label{eq:alphared}\\
\beta_{\omega^{\prime}\omega}&\approx&
-\frac{1}{2\pi\sqrt{\omega\omega^{\prime}}}
e^{-i\omega u_{0}-i\omega^{\prime}v_{0}}
\sum^{\infty}_{k=0}\sum^{k}_{j=0}(-1)^{k-j}i^{k}
\frac{C^{k}_{~j}}{k!}a^{j}\frac{\omega^kc^{(\delta-1)(k-j)}}
{(\omega^{\prime})^{k+(\delta-1)(k-j)}}
\int^{\omega^{\prime}d}_{0}dx e^{ix}x^{j+\delta(k-j)},
\label{eq:betared}
\end{eqnarray}
where we have introduced the frequency $c$ which characterizes
the singularity as
\begin{equation}
c\equiv b^{1/(\delta-1)}.
\label{eq:c}
\end{equation}
They also have another expansion:
\begin{eqnarray}
\alpha_{\omega^{\prime}\omega}&\approx&
\frac{1}{2\pi}\sqrt{\frac{\omega^{\prime}}{\omega}}
e^{i\omega u_{0}-i\omega^{\prime}v_{0}}
\frac{1}{a\omega}
\sum^{\infty}_{k=0}\sum^{k}_{j=0}
i^{k}\frac{C^{k}_{~j}}{k!}
\frac{1}{a^{j+(\delta-1)(k-j)}}
\frac{(\omega^{\prime})^{j}c^{(\delta-1)(k-j)}}
{\omega^{j+(\delta-1)(k-j)}}\nonumber \\
& &\int^{a\omega d}_{0}dx e^{-ix}x^{j+\delta (k-j)}, 
\label{eq:alphaviolet}\\
\beta_{\omega^{\prime}\omega}&\approx&
-\frac{1}{2\pi}\sqrt{\frac{\omega^{\prime}}{\omega}}
e^{-i\omega u_{0}-i\omega^{\prime}v_{0}}
\frac{1}{a\omega}
\sum^{\infty}_{k=0}\sum^{k}_{j=0}
(-1)^{(2\delta-1)(k-j)}i^{k}\frac{C^{k}_{~j}}{k!}
\frac{1}{a^{j+(\delta-1)(k-j)}}
\frac{(\omega^{\prime})^{j}c^{(\delta-1)(k-j)}}
{\omega^{j+(\delta-1)(k-j)}}\nonumber \\
& &\int^{a\omega d}_{0}dx e^{ix}x^{j+\delta (k-j)}.
\label{eq:betaviolet}
\end{eqnarray}
They also have another expansion:
\begin{eqnarray}
\alpha_{\omega^{\prime}\omega}&\approx&
\frac{1}{2\pi}\sqrt{\frac{\omega^{\prime}}{\omega}}
e^{i\omega u_{0}-i\omega^{\prime}v_{0}}
\frac{1}{\delta\omega}
\sum_{k=0}^{\infty}\frac{i^{k}}{k!}
\left(\frac{\omega^{\prime}}{\omega}-a\right)^{k}
\left(\frac{\omega}{c}\right)^{(1+k)(1-\frac{1}{\delta})}
\int^{\frac{\omega}{c}(cd)^{\delta}}_{0}
dx e^{ix}x^{\frac{1+k}{\delta}}, \\
\beta_{\omega^{\prime}\omega}&\approx&
\frac{1}{2\pi}\sqrt{\frac{\omega^{\prime}}{\omega}}
e^{-i\omega u_{0}-i\omega^{\prime}v_{0}}
\frac{1}{\delta\omega}
\sum_{k=0}^{\infty}\frac{i^{k}}{k!}
\left(\frac{\omega^{\prime}}{\omega}+a\right)^{k}
\left(\frac{\omega}{c}\right)^{(1+k)(1-\frac{1}{\delta})}
\int^{\frac{\omega}{c}(cd)^{\delta}}_{0}
dx e^{-ix}x^{\frac{1+k}{\delta}}.
\end{eqnarray}

For $\omega^{\prime}\to 0$, as seen in Eqs.~(\ref{eq:alphared})
and (\ref{eq:betared}), they behave as
\begin{eqnarray}
\alpha_{\omega^{\prime}\omega}&\approx& \frac{i}{2\pi
\sqrt{\omega\omega^{\prime}}}
e^{i\omega u_{0}-i\omega^{\prime}v_{0}},\\
\beta_{\omega^{\prime}\omega}&\approx& -\frac{i}{2\pi
\sqrt{\omega\omega^{\prime}}}
e^{-i\omega u_{0}-i\omega^{\prime}v_{0}}.
\end{eqnarray} 
For $\omega^{\prime}\to \infty$, 
as seen in Eqs.~(\ref{eq:alphaviolet})
and (\ref{eq:betaviolet}),
they behave as
\begin{eqnarray}
\alpha_{\omega^{\prime}\omega}&\approx& \frac{1}{2\pi}
\frac{(\omega^{\prime})^{1/2}}{a\omega^{3/2}}
e^{i\omega u_{0}}\int^{a\omega d}_{0}dx e^{-i x}, \\
\beta_{\omega^{\prime}\omega}&\approx& \frac{1}{2\pi}
\frac{(\omega^{\prime})^{1/2}}{a\omega^{3/2}}
e^{-i\omega u_{0}}\int^{a\omega d}_{0}dx e^{i x},
\end{eqnarray} 
where we have used the formula
\begin{equation}
\int^{\infty}_{0}dx e^{-(\epsilon+i a)x}x^{z-1}=\frac{\Gamma(z)}
{(\epsilon+i a)^{z}},
\end{equation}
for $\epsilon>0$ and $z>0$.

\newpage

\begin{figure}[htbp]
        \centerline{\epsfxsize 18cm \epsfysize 14cm \epsfbox{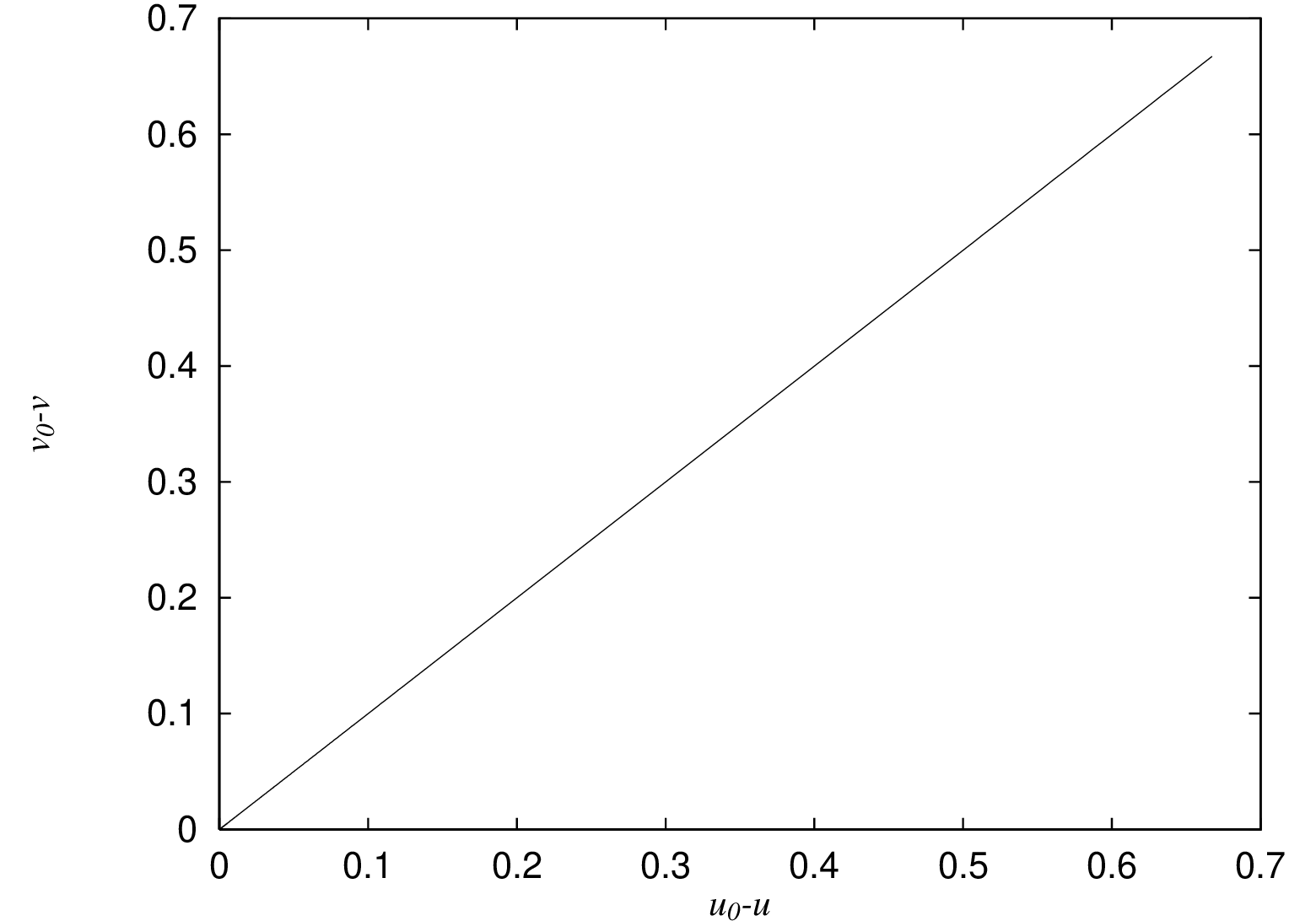}}
~(a)
\end{figure}
\begin{figure}
        \centerline{\epsfxsize 18cm \epsfysize 14cm \epsfbox{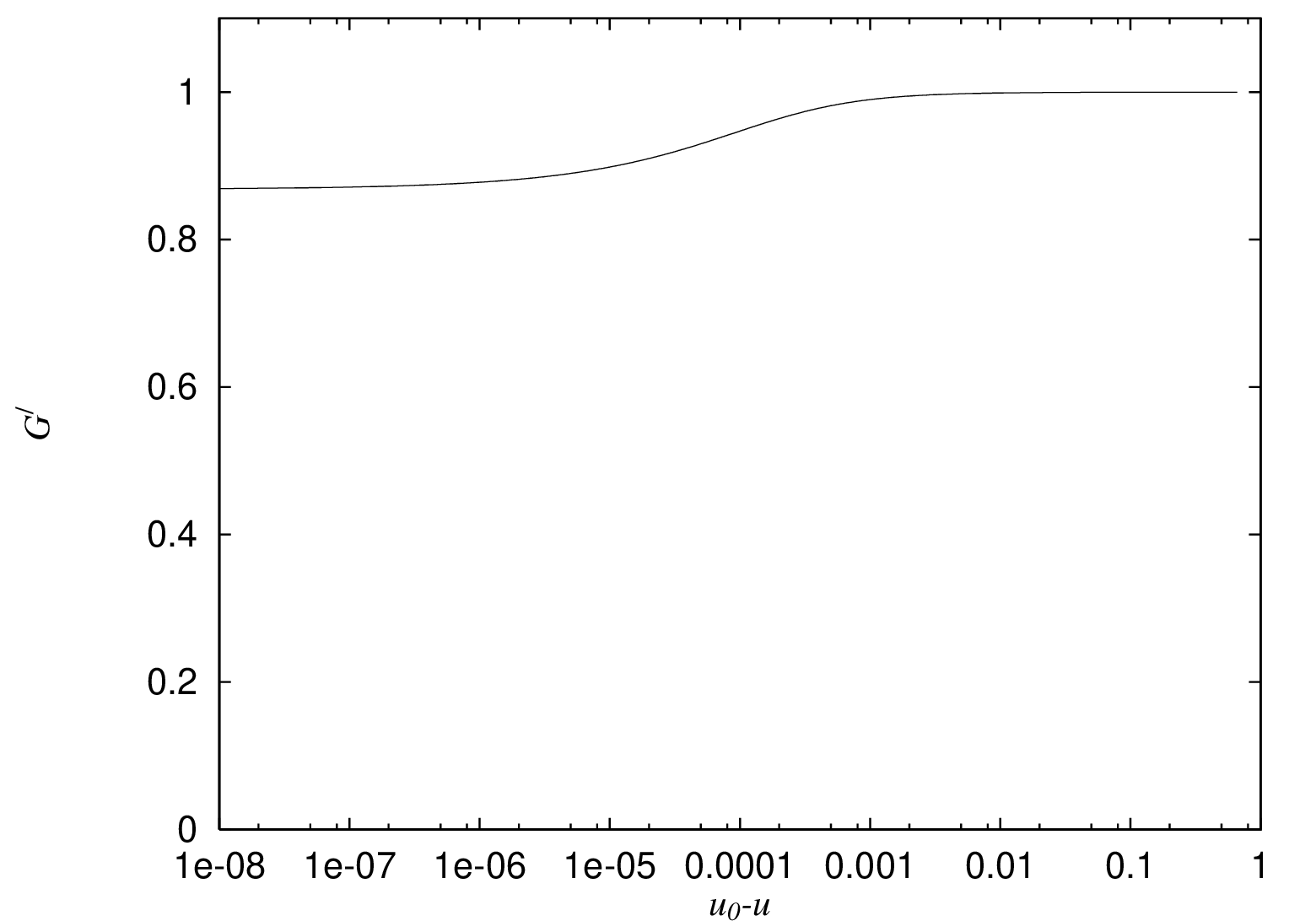}}
~(b)
\end{figure}
\begin{figure}[htbp]
 \centerline{\epsfxsize 18cm \epsfysize 14cm \epsfbox{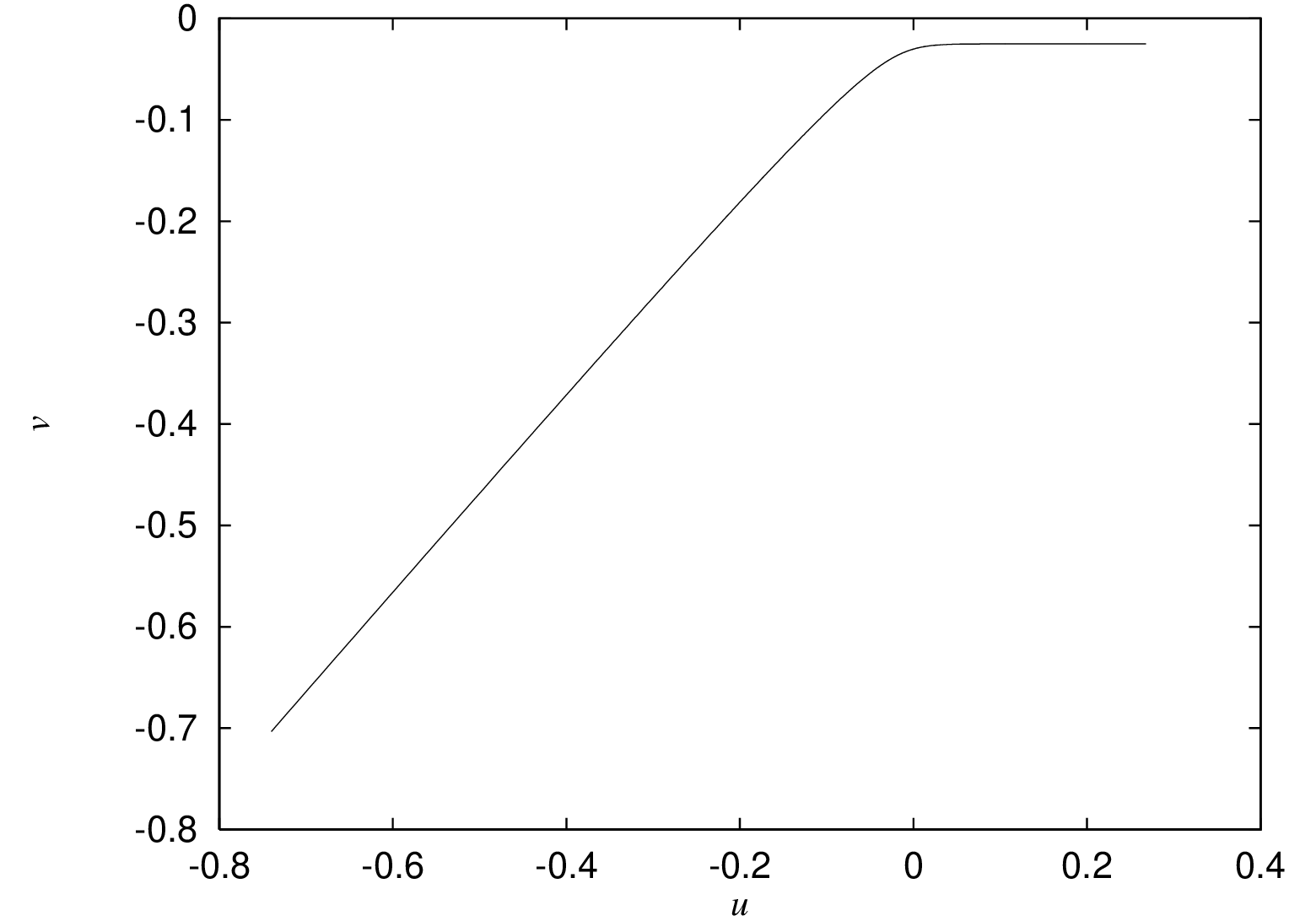}}
 ~(a)
\end{figure}
\begin{figure}
 \centerline{\epsfxsize 18cm \epsfysize 14cm \epsfbox{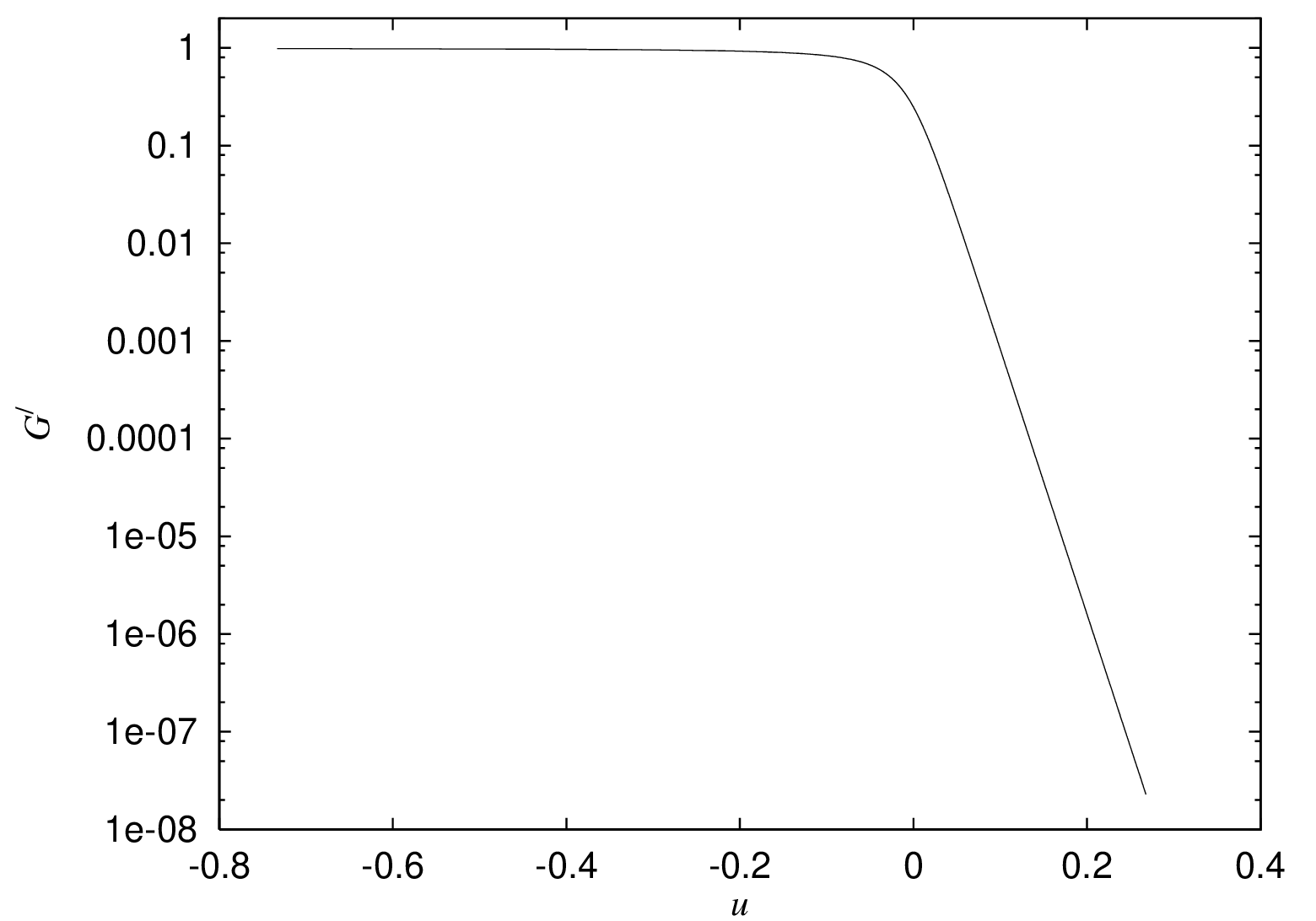}}
 ~(b)
\end{figure}

\end{document}